\documentclass[aps,twocolumn,prl,superscriptaddress]{revtex4-2}

\usepackage[dvipdfmx]{graphicx}
\usepackage{subfigure}
\usepackage{multirow}

\usepackage{fancyhdr}
\usepackage{longtable}
\usepackage{parskip}
\usepackage[T1]{fontenc}
\usepackage{dcolumn}   

\usepackage{bm}        
\usepackage{amsfonts}  
\usepackage{amsmath}   
\usepackage{amssymb}   
\usepackage{xspace}

\newcommand{\tns}{Ta$_2$NiSe$_5$\xspace}
\newcommand{\raa}{$aa$\xspace}
\newcommand{\rac}{$ac$\xspace}
\newcommand{\ag}{$A_{g}$\xspace}
\newcommand{\bg}{$B_{2g}$\xspace}

\newcommand{\tpp}{$t_{pp}$\xspace}
\newcommand{\tc}{$T_{\text{c}}$\xspace}

\newcommand{\mjc}{mJ/cm$^2$\xspace}
\newcommand{\wn}{cm$^{-1}$\xspace}

\begin{document}
	\title{Disentangling lattice and electronic instabilities in the excitonic insulator candidate \tns by nonequilibrium spectroscopy}
	\author{Kota~Katsumi}
	\affiliation{Université Paris Cité, Matériaux et Phénomènes Quantiques UMR CNRQ 7162, Bâtiment Condorcet, 75205 Paris Cedex 13, France}
	\author{Alexandr~Alekhin}
	\affiliation{Université Paris Cité, Matériaux et Phénomènes Quantiques UMR CNRQ 7162, Bâtiment Condorcet, 75205 Paris Cedex 13, France}
	\author{Sofia-Michaela~Souliou}
	\affiliation{Institute for Quantum Materials and Technologies, Karlsruhe Institute of Technology, 76021 Karlsruhe, Germany}
	\author{Michael Merz}
	\affiliation{Institute for Quantum Materials and Technologies, Karlsruhe Institute of Technology, 76021 Karlsruhe, Germany}
	\affiliation{Karlsruhe Nano Micro Facility (KNMFi), Karlsruhe Institute of Technology, 76344 Eggenstein-Leopoldshafen, Germany}
	\author{Amir-Abbas~Haghighirad}
	\affiliation{Institute for Quantum Materials and Technologies, Karlsruhe Institute of Technology, 76021 Karlsruhe, Germany}
	\author{Matthieu~Le~Tacon}
	\affiliation{Institute for Quantum Materials and Technologies, Karlsruhe Institute of Technology, 76021 Karlsruhe, Germany}
	\author{Sarah~Houver}
	\affiliation{Université Paris Cité, Matériaux et Phénomènes Quantiques UMR CNRQ 7162, Bâtiment Condorcet, 75205 Paris Cedex 13, France}
	\author{Maximilien~Cazayous}
	\affiliation{Université Paris Cité, Matériaux et Phénomènes Quantiques UMR CNRQ 7162, Bâtiment Condorcet, 75205 Paris Cedex 13, France}
	\author{Alain~Sacuto}
	\affiliation{Université Paris Cité, Matériaux et Phénomènes Quantiques UMR CNRQ 7162, Bâtiment Condorcet, 75205 Paris Cedex 13, France}
	\author{Yann~Gallais}
	\affiliation{Université Paris Cité, Matériaux et Phénomènes Quantiques UMR CNRQ 7162, Bâtiment Condorcet, 75205 Paris Cedex 13, France}
	
	\begin{abstract}
		\tns is an excitonic insulator candidate showing the semiconductor/semimetal-to-insulator (SI) transition below \tc = 326~K. However, since a structural transition accompanies the SI transition, deciphering the role of electronic and lattice degrees of freedom in driving the SI transition has remained controversial. Here, we investigate the photoexcited nonequilibrium state in \tns using pump-probe Raman and photoluminescence (PL) spectroscopies. The combined nonequilibrium spectroscopic measurements of the lattice and electronic states reveal the presence of a photoexcited metastable state where the insulating gap is suppressed, but the low-temperature structural distortion is preserved. We conclude that electron correlations play a vital role in the SI transition of \tns.
	\end{abstract}
	
	\maketitle
	Many-body interactions in quantum materials are one of the main drivers of exotic ground states, such as Mott insulators, superconductors, and density waves. In narrow-gap semiconductors or semimetals, the Coulomb interaction between electrons and holes may lead to a spontaneous formation of excitons. These excitons are expected to condense and give rise to an unconventional insulating ground state called excitonic insulator \cite{mott_transition_1961, cloizeaux_exciton_1965,Jerome1967,Kohn1967,Halperin1968}. Among the various excitonic insulator candidate materials, \tns (TNS) is a prototypical example because it has a small, possibly vanishing, direct band gap \cite{Wakisaka2009,Wakisaka2012} and has no instability at finite wave-vector such as charge density wave order. With decreasing temperature, TNS exhibits a semiconductor/semimetal-to-insulator (SI) transition below the transition temperature \tc~=~326~K \cite{Salvo1986,Lu2017}, which has been associated with an excitonic insulator transition due to electronic correlations. This scenario is supported by the band flattening observed in angle-resolved photoemission spectroscopy (ARPES) \cite{Wakisaka2009,Wakisaka2012,Seki2014,Fukutani2021}, and the opening of a large gap in optical spectra \cite{Lu2017,Larkin2017,Larkin2018,Sugimoto2018}. Furthermore, the key role of electronic correlations in driving SI transition is supported by various theoretical studies \cite{Kaneko2013,Ejima2014,Sugimoto2016,Yamada2016,Sugimoto2018}, and by the observation of critical charge fluctuations near the SI transition by Raman scattering \cite{Kim2021,Volkov2021_npj,Ye2021}.
	
	However, assessing the origin of the SI transition is complicated by the accompanying structural transition from a high-temperature orthorhombic phase to low-temperature monoclinic phase \cite{Salvo1986}. It has been argued that the monoclinic structural distortion induces the hybridization of the conduction and valence bands, resulting in the insulator gap opening \cite{Nakano2018,Mazza2020,Watson2020}. In this scenario, the lattice degrees of freedom play a decisive, possibly dominant, role in the phase transition \cite{zenker_fate_2014,Subedi2020, Windgatter2021}. To date, the respective roles of electronic and lattice degrees of freedom in the SI phase transition of TNS remain an open question.
	
	To investigate the origin of the SI transition in TNS, pump-probe spectroscopy is a promising technique because it can in principle, resolve the origin of the insulating gap using the inherently different timescales of the electronic and the lattice degrees of freedom \cite{hellmann_time-domain_2012,Gianetti2016,Lloyd_Hughes_2021}. For TNS, previous pump-probe experiments focused on the photoexcited dynamics of the insulating gap using time-resolved reflectivity \cite{Mor2018,Werdehausen2018,Bretscher2021,Liu2021,Miyamoto2022} and Angle-Resolved Photo-Emission Spectroscopy (ARPES) \cite{Mor2017,Okazaki2018,Tang2020,Baldini2020,Suzuki2021,Golez2021,Saha2021,Mor2022}. However, these measurements have yielded conflicting results, and how the nonequilibrium dynamics of the insulating gap correlates with the monoclinic distortion has not been clarified. Thus, a nonequilibrium approach that can simultaneously probe the insulating gap and the monoclinic distortion is desirable to resolve the origin of the SI transition of TNS.
	
	In this letter, we report pump-probe Raman scattering and photoluminescence (PL) experiments performed with the aim of investigating both the structural symmetry breaking and the insulating gap dynamics of TNS following a destabilizing sub-picosecond optical pulse. Polarization-resolved transient Raman phonon spectra unambiguously demonstrate that the structural symmetry remains in the low-temperature monoclinic phase a few picoseconds after photoexcitation. In the electronic sector, we show that the SI transition is marked by the emergence of an emission peak which is activated by the opening of the insulating gap below \tc. In the pump-probe spectrum, the PL emission intensity is suppressed to its equilibrium value at 300~K, indicating a photo-induced gap closing. Remarkably, the quenched emission signal persists at least 18~ps after the pump, signifying a metastable state with a collapsed gap but a robust monoclinic distortion. Our results indicate that the monoclinic distortion plays a marginal role in the opening of the insulating gap, leading us to conclude that the SI transition in TNS is electronic-driven.
	
	\begin{figure}[t]
		\centering
		\includegraphics[width=\columnwidth]{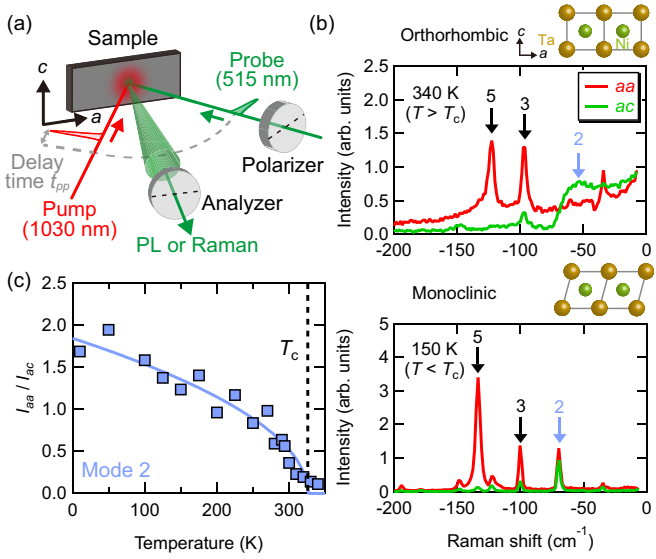}
		\caption{(a) Schematic illustration of the pump-probe polarization-resolved Raman/PL measurements. (b) The anti-Stokes Raman spectra at 340~K (top) and 150~K (bottom) measured with a 514.5-nm CW laser. The black and light blue vertical arrows denote the \ag and \bg modes defined above \tc, respectively. (c) The mode-2 phonon intensity ratio between the \raa and \rac geometry $I_{aa}/I_{ac}$ as a function of temperature. The solid curve is the fitting curve. The black vertical dashed line denotes \tc.}
		\label{fig1}
	\end{figure}
	
	The schematic of the pump-probe Raman and PL experiments is illustrated in Fig.~\ref{fig1}(a). The 1030-nm pump and 515-nm probe pulses are polarized along the $a$-axis of the sample. The analyzer sets the polarization of the measured Raman or PL spectra to the $a$-axis (\raa geometry) or $c$-axis (\rac geometry). For the pump-probe Raman measurements, we tailored the probe pulse at 514.5~nm with a spectral width of 9~\wn, resulting in a time resolution of 2.5~ps, while that in the pump-probe PL measurements was 700~fs. The details of the sample fabrication and experimental method are described in the Supplemental material (SM) \cite{sm}. Complementary equilibrium Raman and PL measurements were performed using the emission lines of an Ar/Kr laser and several solid-state lasers (see SM). 
	
	In Fig. \ref{fig1}(b), we present the equilibrium anti-Stokes Raman spectra using a continuous wave (CW) 514.5-nm laser. Here the choice to display the anti-Stokes spectrum is motivated by the comparison with the time-resolved Raman data discussed below (see also \cite{sm}). In the high-temperature orthorhombic phase at 340~K (the upper panel), the \rac and \raa geometries probe phonons with \bg and \ag representations, respectively (see also \cite{sm} for the complete temperature dependence of the Raman spectra and a discussion of Raman phonon selection rules in TNS). The equilibrium Raman phonon spectra agree with previously published results \cite{Kim2020,Kim2021,Volkov2021_npj}. Hereafter, we will focus on three dominant phonon modes labeled 2, 3, and 5 following previous notations \cite{Kim2021}). Of particular interest is the mode-2 phonon which corresponds to a monoclinic shearing motion of the Ta atoms and displays the most evident fingerprints of the structural transition. Above \tc, the mode-2 phonon is observed only in the \rac geometry as expected for a \bg phonon. It exhibits a significant broadening which was previously ascribed to a strong electron-phonon coupling \cite{Kim2021,Volkov2021_npj,Ye2021}. Below \tc at 150~K (lower panel of Fig. \ref{fig1}(b)), the mode-2 phonon sharpens and becomes activated in the \raa spectra because the \ag and \bg representations are no longer distinct in the low-temperature monoclinic phase \cite{Hayes2012,Kim2021,sm}. Figure \ref{fig1}(c) plots the mode-2 phonon intensity ratio between the \raa and \rac geometry $I_{aa}/I_{ac}$ as a function of temperature. The ratio $I_{aa}/I_{ac}$ displays an increase below \tc and is well fitted by $\alpha\sqrt{1-T/T_{\text{c}}}$ ($\alpha$ is the fitting parameter), demonstrating that it serves as a reporter of the monoclinic distortion below \tc.
	
	\begin{figure}[t]
		\centering
		\includegraphics[width=\columnwidth]{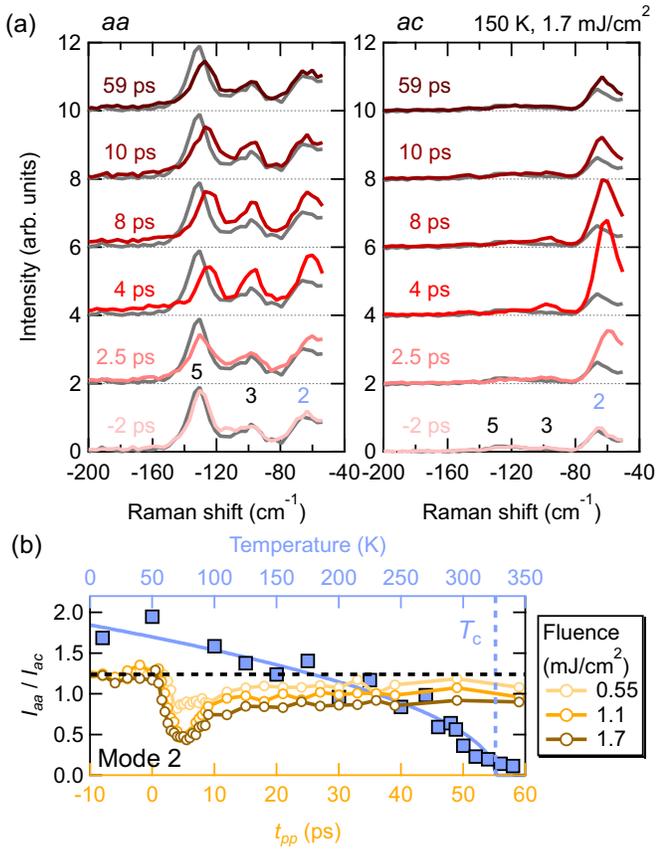}
		\caption{(a) The transient anti-Stokes Raman spectra at 150~K for the \raa (left) and \rac (right) geometries at selected delay times \tpp for the pump fluence of 1.7~\mjc. The gray curves show the probe-only anti-Stokes Raman spectra at 150~K. The mode numbers are denoted by black (mode 3 and 5) and light blue (mode 2) for the \ag and \bg phonons defined above \tc, respectively. (b) The mode-2 phonon intensity ratio between the \raa and \rac geometry $I_{aa}/I_{ac}$ as a function of \tpp (orange curves) for selected pump fluences. The light blue data points and curve are the equilibrium temperature-dependent Raman data shown in Fig.~\ref{fig1}(c). The black horizontal dashed line denotes $I_{aa}/I_{ac}$ at 150~K in equilibrium.}
		\label{fig2}
	\end{figure}
	
	Having established that the mode-2 phonon probes the structural symmetry breaking below \tc, we now describe the pump-probe Raman results. All the pump-probe measurements were performed at the spot temperature of 150~K, estimated using the probe-only Stokes-anti-Stokes relation for the mode-5 phonon \cite{sm,Hayes2012}. Figure \ref{fig2}(a) shows the transient anti-Stokes spectra with the incoming pump fluence of 1.7~\mjc measured at selected pump-probe delay time \tpp. As described in SM, the anti-Stokes spectra above $-55$ ~\wn are blocked by the edge filter and not shown \cite{sm}. Corresponding Stokes spectra were also measured, but with a higher energy cut-off which hinders the mode-2 phonon \cite{sm}. Upon photoexcitation for the \raa geometry, the mode-2 and mode-3 phonon intensities increase, whereas the mode-5 phonon intensity decreases. As discussed in the SM, these behaviors result from a combination of pump-induced increases in the phonon population and pump-induced changes in the phonon Raman response \cite{sm}. 
	\begin{figure}[t]
		\centering
		\includegraphics[width=\columnwidth]{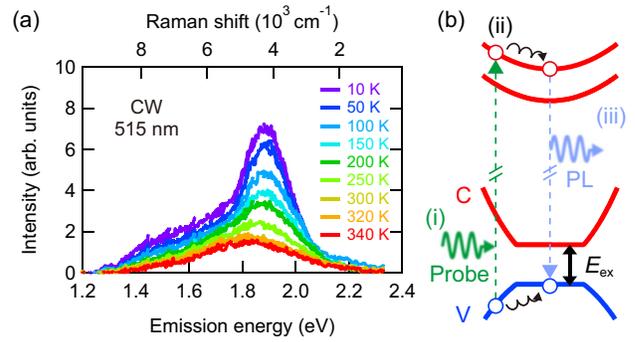}
		\caption{(a) High energy Raman/PL spectra measured with CW laser (514.5~nm) in the \raa geometry at selected temperatures as a function of Raman shift (top axis) and photon emission energy (bottom axis). (b) Schematic illustration of the PL mechanism. $E_{ex}$ denotes the insulating-gap energy.}
		\label{fig3}
	\end{figure}
	
	The mode-2 phonon intensity ratio $I_{aa}/I_{ac}$ as a function of \tpp is summarized in Fig.~\ref{fig2}(b). For the largest fluence, a significant weakening of the monoclinic distortion during the first few picoseconds is observed. It is followed by a recovery to a metastable value close to the equilibrium value at 150~K after 15~ps. Interestingly, for all the measured delay times and all the pump fluences, $I_{aa}/I_{ac}$ remains finite, indicating a preserved monoclinic distortion. While we cannot discard the possibility that the structural symmetry switches to the high-temperature orthorhombic phase during the first few picoseconds due to the limited time resolution, this result evidences that at least after $t_{pp}$~=~5~ps, the structural symmetry remains in the low-temperature monoclinic phase at all fluences.
	
	
	Next, we investigate the spectra at higher Raman shifts. Figure \ref{fig3}(a) shows the temperature-dependence of the equilibrium Raman spectra using a CW laser wavelength of $\lambda_0 = 514.5~\text{nm}$. Above \tc, the spectrum displays a broad peak that gains considerable intensity upon cooling below \tc. As shown in Fig.~S2(a) of the SM, the position of the peak, when plotted in terms of Raman shift, significantly depends on the CW laser wavelength $\lambda_0$. Instead, it coincides when plotted as a function of the emitted photon energy, demonstrating that the observed peak is not a Raman but a PL process, where the photoexcited electron-hole pairs first relax before recombining by emitting a photon, as schematically shown in Fig. \ref{fig3}(b). We note that our measurement and its associated interpretation contradict an earlier report where the peak in the insulating phase was assigned to an electronic Raman process across the insulating gap \cite{Volkov2021_npj}. The peak emission energy, 1.9~eV, is significantly larger than the insulating-gap energy $E_{ex}\simeq$~160~meV reported in previous optical spectroscopy measurements \cite{Lu2017,Larkin2017,Larkin2018,Sugimoto2018}, indicating that the PL process does not correspond to recombination across the insulating gap. We assign it to non-thermalized "hot" PL \cite{Yu} where excited high-energy conduction band electrons recombine with holes in the top-most valence band before fully thermalizing (or equivalently, low-energy valence band holes recombine with electrons in the lowest conduction band, as depicted in Fig.~\ref{fig3}(b)). The enhanced PL emission below \tc is assigned to the opening of the insulating gap, which quenches non-radiative recombination processes involving phonon emission, thus significantly increasing the efficiency of PL radiative processes \cite{Fox2010}.
	
	\begin{figure}[t]
		\centering
		\includegraphics[width=\columnwidth]{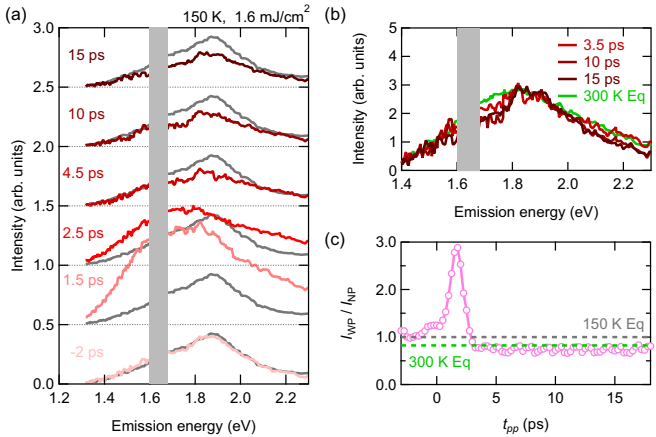}
		\caption{(a) Pump-probe PL spectra at 150~K for the \raa geometry at selected delay times \tpp and for the pump fluence of 1.6~\mjc. The gray curves show the probe-only PL spectra at 150~K. The gray-shaded spectral region is blocked by the notch filter used to remove stray light. (b) Expanded figure of (a) at selected delay times. The green curve shows the probe-only PL spectra at 300~K. (c) The time evolution of the integrated transient PL intensity from 1.3~eV to 2.3~eV ($I_{\text{WP}}$) divided by that without photoexcitation ($I_{\text{NP}}$). The horizontal dashed lines denote the integrated PL intensity without the pump at 150~K (gray) and 300~K (green).}
		\label{fig4}
	\end{figure}
	Transient PL spectra at 150~K with the incident pump fluence of 1.6~\mjc are shown in Fig.~\ref{fig4}(a). Right after the pump pulse (\tpp~$<$~2~ps), the PL intensity shows an overall increase, while it decreases after \tpp~=~2~ps. We attribute the initial enhancement of the PL intensity to pump-induced excess holes/electrons in the valence/conduction band, which increase the occupancy factors governing the PL process \cite{Fox2010}. It could also be linked to a pump-induced transient increase of the insulating gap as reported in the previous pump-probe ARPES measurements \cite{Mor2017,Golez2021}. Remarkably, the initial increase is followed by a substantial reduction of the PL intensity, which sustains up to 18~ps. Fig.~\ref{fig4}(b) shows that the PL spectra at selected delays after 3.5~ps is almost identical to the probe-only spectrum at 300~K. This is further illustrated by the transient PL integrated area, which falls slightly below that of the 300~K equilibrium spectra over the whole time window, with hardly any time dependence from 3.5~ps to at least 18~ps (Fig. \ref{fig4}(c)), suggesting a metastable state with a collapsed insulating gap. Crucially, since the pump-probe Raman results displayed in Fig. \ref{fig2}(b) demonstrate that the crystal remains in the low-temperature monoclinic phase over the same time window, we can conclude that the monoclinic structural distortion plays a marginal role in the insulating-gap formation below \tc. Our results thus support a leading role for electron correlations in establishing the insulating phase.
	
	We now discuss the possible origins of the observed metastable state. Within an excitonic insulator picture, the metastable state can be ascribed to the photoexcited carriers, which screen the Coulomb interaction and prevent the reformation of excitons as proposed in \cite{Mor2017,Mor2022}. Another possibility is a structural change, distinct from the monoclinic one, which would significantly weaken the electronic instability responsible for the insulating gap. Indeed, recent mean-field calculations with a realistic electronic band structure of TNS suggest that an electronically driven insulating phase is only realized in a rather narrow parameter space of on-site and inter-site Coulomb interactions, thus making it highly susceptible to even small changes in the lattice structure \cite{Mazza2020,Windgatter2021}.
	
	In fact, fingerprints of metastable structural changes are observed in the transient Raman phonon spectra. In particular, the intensity and peak position of the mode-5 phonon do not relax to the values in equilibrium even at \tpp~=~59~ps, as shown in Fig. \ref{fig2}(a). The shift in the mode-5 phonon energy presented in Fig. S5(c) is sizable ($\sim$~3~\wn at \tpp~=~59~ps) and cannot be ascribed to a transient heating effect since the temperature of the mode-5 phonon relaxes within 10~ps (Fig. S5(g)). We note that the mode-5 phonon belongs to the symmetry preserving the \ag representation in the orthorhombic phase. It involves mostly Se and Ni atomic motions perpendicular to the Ta/Ni chain direction and not the shearing motion of the Ta atoms responsible for the monoclinic phase \cite{Suzuki2021}. The observed phonon shift may be indicative of a slight lattice distortion that stabilizes the observed gapless metastable state. Time-resolved X-ray or electron diffraction measurements using a similar pump fluence regime are desirable to confirm this scenario.
	
	In summary, our combined pump-probe Raman and PL spectroscopic measurements on TNS reveal the presence of a long-lived metastable state with a collapsed insulating gap, but a preserved monoclinic lattice distortion. These results lead us to conclude that the insulating-gap formation cannot be explained solely by the structural distortion, and hence that electron correlations play a leading role. The transient Raman phonon data suggest that a pump-induced lattice distortion, distinct from the monoclinic one, might be responsible for the pump-induced metastable state.
	
	We acknowledge Alaska Subedi for fruitful discussions.
	ANR (Agence Nationale de la Recherche) and CGI (Commissariat G\'{e}n\'{e}ral \`{a} l'Investissement) are acknowledged for their financial support of this work through Labex SEAM (Science and Engineering for Advanced Materials and devices), ANR-10-LABX-0096 and ANR-18-IDEX-0001. This work was also supported by a DIM SIRTEQ grant from Ile-de-France region and an Emergence grant from IDEX Université Paris Cité.

	\bibliography{TNS.bib}

\begin{thebibliography}{48}%
\makeatletter
\providecommand \@ifxundefined [1]{%
 \@ifx{#1\undefined}
}%
\providecommand \@ifnum [1]{%
 \ifnum #1\expandafter \@firstoftwo
 \else \expandafter \@secondoftwo
 \fi
}%
\providecommand \@ifx [1]{%
 \ifx #1\expandafter \@firstoftwo
 \else \expandafter \@secondoftwo
 \fi
}%
\providecommand \natexlab [1]{#1}%
\providecommand \enquote  [1]{``#1''}%
\providecommand \bibnamefont  [1]{#1}%
\providecommand \bibfnamefont [1]{#1}%
\providecommand \citenamefont [1]{#1}%
\providecommand \href@noop [0]{\@secondoftwo}%
\providecommand \href [0]{\begingroup \@sanitize@url \@href}%
\providecommand \@href[1]{\@@startlink{#1}\@@href}%
\providecommand \@@href[1]{\endgroup#1\@@endlink}%
\providecommand \@sanitize@url [0]{\catcode `\\12\catcode `\$12\catcode
  `\&12\catcode `\#12\catcode `\^12\catcode `\_12\catcode `\%12\relax}%
\providecommand \@@startlink[1]{}%
\providecommand \@@endlink[0]{}%
\providecommand \url  [0]{\begingroup\@sanitize@url \@url }%
\providecommand \@url [1]{\endgroup\@href {#1}{\urlprefix }}%
\providecommand \urlprefix  [0]{URL }%
\providecommand \Eprint [0]{\href }%
\providecommand \doibase [0]{http://dx.doi.org/}%
\providecommand \selectlanguage [0]{\@gobble}%
\providecommand \bibinfo  [0]{\@secondoftwo}%
\providecommand \bibfield  [0]{\@secondoftwo}%
\providecommand \translation [1]{[#1]}%
\providecommand \BibitemOpen [0]{}%
\providecommand \bibitemStop [0]{}%
\providecommand \bibitemNoStop [0]{.\EOS\space}%
\providecommand \EOS [0]{\spacefactor3000\relax}%
\providecommand \BibitemShut  [1]{\csname bibitem#1\endcsname}%
\let\auto@bib@innerbib\@empty
\bibitem [{\citenamefont {Mott}(1961)}]{mott_transition_1961}%
  \BibitemOpen
  \bibfield  {author} {\bibinfo {author} {\bibfnamefont {N.~F.}\ \bibnamefont
  {Mott}},\ }\href@noop {} {\bibfield  {journal} {\bibinfo  {journal} {The
  Philosophical Magazine: A Journal of Theoretical Experimental and Applied
  Physics}\ }\textbf {\bibinfo {volume} {6}},\ \bibinfo {pages} {287} (\bibinfo
  {year} {1961})}\BibitemShut {NoStop}%
\bibitem [{\citenamefont {Cloizeaux}(1965)}]{cloizeaux_exciton_1965}%
  \BibitemOpen
  \bibfield  {author} {\bibinfo {author} {\bibfnamefont {J.~D.}\ \bibnamefont
  {Cloizeaux}},\ }\href@noop {} {\bibfield  {journal} {\bibinfo  {journal} {J.
  Phys. Chem. Solids}\ }\textbf {\bibinfo {volume} {26}},\ \bibinfo {pages}
  {259} (\bibinfo {year} {1965})}\BibitemShut {NoStop}%
\bibitem [{\citenamefont {J\'{e}rome}\ \emph {et~al.}(1967)\citenamefont
  {J\'{e}rome}, \citenamefont {Rice},\ and\ \citenamefont {Kohn}}]{Jerome1967}%
  \BibitemOpen
  \bibfield  {author} {\bibinfo {author} {\bibfnamefont {D.}~\bibnamefont
  {J\'{e}rome}}, \bibinfo {author} {\bibfnamefont {T.~M.}\ \bibnamefont
  {Rice}}, \ and\ \bibinfo {author} {\bibfnamefont {W.}~\bibnamefont {Kohn}},\
  }\href@noop {} {\bibfield  {journal} {\bibinfo  {journal} {Phys. Rev.}\
  }\textbf {\bibinfo {volume} {158}},\ \bibinfo {pages} {462} (\bibinfo {year}
  {1967})}\BibitemShut {NoStop}%
\bibitem [{\citenamefont {Kohn}(1967)}]{Kohn1967}%
  \BibitemOpen
  \bibfield  {author} {\bibinfo {author} {\bibfnamefont {W.}~\bibnamefont
  {Kohn}},\ }\href@noop {} {\bibfield  {journal} {\bibinfo  {journal} {Phys.
  Rev. Lett.}\ }\textbf {\bibinfo {volume} {19}},\ \bibinfo {pages} {439}
  (\bibinfo {year} {1967})}\BibitemShut {NoStop}%
\bibitem [{\citenamefont {Halperin}\ and\ \citenamefont
  {Rice}(1968)}]{Halperin1968}%
  \BibitemOpen
  \bibfield  {author} {\bibinfo {author} {\bibfnamefont {B.~I.}\ \bibnamefont
  {Halperin}}\ and\ \bibinfo {author} {\bibfnamefont {T.~M.}\ \bibnamefont
  {Rice}},\ }\href@noop {} {\bibfield  {journal} {\bibinfo  {journal} {Rev.
  Mod. Phys.}\ }\textbf {\bibinfo {volume} {40}},\ \bibinfo {pages} {755}
  (\bibinfo {year} {1968})}\BibitemShut {NoStop}%
\bibitem [{\citenamefont {Wakisaka}\ \emph {et~al.}(2009)\citenamefont
  {Wakisaka}, \citenamefont {Sudayama}, \citenamefont {Takubo}, \citenamefont
  {Mizokawa}, \citenamefont {Arita}, \citenamefont {Namatame}, \citenamefont
  {Taniguchi}, \citenamefont {Katayama}, \citenamefont {Nohara},\ and\
  \citenamefont {Takagi}}]{Wakisaka2009}%
  \BibitemOpen
  \bibfield  {author} {\bibinfo {author} {\bibfnamefont {Y.}~\bibnamefont
  {Wakisaka}}, \bibinfo {author} {\bibfnamefont {T.}~\bibnamefont {Sudayama}},
  \bibinfo {author} {\bibfnamefont {K.}~\bibnamefont {Takubo}}, \bibinfo
  {author} {\bibfnamefont {T.}~\bibnamefont {Mizokawa}}, \bibinfo {author}
  {\bibfnamefont {M.}~\bibnamefont {Arita}}, \bibinfo {author} {\bibfnamefont
  {H.}~\bibnamefont {Namatame}}, \bibinfo {author} {\bibfnamefont
  {M.}~\bibnamefont {Taniguchi}}, \bibinfo {author} {\bibfnamefont
  {N.}~\bibnamefont {Katayama}}, \bibinfo {author} {\bibfnamefont
  {M.}~\bibnamefont {Nohara}}, \ and\ \bibinfo {author} {\bibfnamefont
  {H.}~\bibnamefont {Takagi}},\ }\href@noop {} {\bibfield  {journal} {\bibinfo
  {journal} {Phys. Rev. Lett.}\ }\textbf {\bibinfo {volume} {103}},\ \bibinfo
  {pages} {026402} (\bibinfo {year} {2009})}\BibitemShut {NoStop}%
\bibitem [{\citenamefont {Wakisaka}\ \emph {et~al.}(2012)\citenamefont
  {Wakisaka}, \citenamefont {Sudayama}, \citenamefont {Takubo}, \citenamefont
  {Mizokawa}, \citenamefont {Saini}, \citenamefont {Arita}, \citenamefont
  {Namatame}, \citenamefont {Taniguchi}, \citenamefont {Katayama},
  \citenamefont {Nohara},\ and\ \citenamefont {Takagi}}]{Wakisaka2012}%
  \BibitemOpen
  \bibfield  {author} {\bibinfo {author} {\bibfnamefont {Y.}~\bibnamefont
  {Wakisaka}}, \bibinfo {author} {\bibfnamefont {T.}~\bibnamefont {Sudayama}},
  \bibinfo {author} {\bibfnamefont {K.}~\bibnamefont {Takubo}}, \bibinfo
  {author} {\bibfnamefont {T.}~\bibnamefont {Mizokawa}}, \bibinfo {author}
  {\bibfnamefont {N.~L.}\ \bibnamefont {Saini}}, \bibinfo {author}
  {\bibfnamefont {M.}~\bibnamefont {Arita}}, \bibinfo {author} {\bibfnamefont
  {H.}~\bibnamefont {Namatame}}, \bibinfo {author} {\bibfnamefont
  {M.}~\bibnamefont {Taniguchi}}, \bibinfo {author} {\bibfnamefont
  {N.}~\bibnamefont {Katayama}}, \bibinfo {author} {\bibfnamefont
  {M.}~\bibnamefont {Nohara}}, \ and\ \bibinfo {author} {\bibfnamefont
  {H.}~\bibnamefont {Takagi}},\ }\href@noop {} {\bibfield  {journal} {\bibinfo
  {journal} {J. Supercond. Nov. Magn.}\ }\textbf {\bibinfo {volume} {25}},\
  \bibinfo {pages} {1231} (\bibinfo {year} {2012})}\BibitemShut {NoStop}%
\bibitem [{\citenamefont {Di~Salvo}\ \emph {et~al.}(1986)\citenamefont
  {Di~Salvo}, \citenamefont {Chen}, \citenamefont {Fleming}, \citenamefont
  {Waszczak}, \citenamefont {Dunn}, \citenamefont {Sunshine},\ and\
  \citenamefont {Ibers}}]{Salvo1986}%
  \BibitemOpen
  \bibfield  {author} {\bibinfo {author} {\bibfnamefont {F.~J.}\ \bibnamefont
  {Di~Salvo}}, \bibinfo {author} {\bibfnamefont {C.~H.}\ \bibnamefont {Chen}},
  \bibinfo {author} {\bibfnamefont {R.~M.}\ \bibnamefont {Fleming}}, \bibinfo
  {author} {\bibfnamefont {J.~V.}\ \bibnamefont {Waszczak}}, \bibinfo {author}
  {\bibfnamefont {R.~G.}\ \bibnamefont {Dunn}}, \bibinfo {author}
  {\bibfnamefont {S.~A.}\ \bibnamefont {Sunshine}}, \ and\ \bibinfo {author}
  {\bibfnamefont {J.~A.}\ \bibnamefont {Ibers}},\ }\href@noop {} {\bibfield
  {journal} {\bibinfo  {journal} {J. less-common met.}\ }\textbf {\bibinfo
  {volume} {116}},\ \bibinfo {pages} {51} (\bibinfo {year} {1986})}\BibitemShut
  {NoStop}%
\bibitem [{\citenamefont {Lu}\ \emph {et~al.}(2017)\citenamefont {Lu},
  \citenamefont {Kono}, \citenamefont {Larkin}, \citenamefont {Rost},
  \citenamefont {Takayama}, \citenamefont {Boris}, \citenamefont {Keimer},\
  and\ \citenamefont {Takagi}}]{Lu2017}%
  \BibitemOpen
  \bibfield  {author} {\bibinfo {author} {\bibfnamefont {Y.~F.}\ \bibnamefont
  {Lu}}, \bibinfo {author} {\bibfnamefont {H.}~\bibnamefont {Kono}}, \bibinfo
  {author} {\bibfnamefont {T.~I.}\ \bibnamefont {Larkin}}, \bibinfo {author}
  {\bibfnamefont {A.~W.}\ \bibnamefont {Rost}}, \bibinfo {author}
  {\bibfnamefont {T.}~\bibnamefont {Takayama}}, \bibinfo {author}
  {\bibfnamefont {A.~V.}\ \bibnamefont {Boris}}, \bibinfo {author}
  {\bibfnamefont {B.}~\bibnamefont {Keimer}}, \ and\ \bibinfo {author}
  {\bibfnamefont {H.}~\bibnamefont {Takagi}},\ }\href@noop {} {\bibfield
  {journal} {\bibinfo  {journal} {Nat. Commun.}\ }\textbf {\bibinfo {volume}
  {8}},\ \bibinfo {pages} {14408} (\bibinfo {year} {2017})}\BibitemShut
  {NoStop}%
\bibitem [{\citenamefont {Seki}\ \emph {et~al.}(2014)\citenamefont {Seki},
  \citenamefont {Wakisaka}, \citenamefont {Kaneko}, \citenamefont {Toriyama},
  \citenamefont {Konishi}, \citenamefont {Sudayama}, \citenamefont {Saini},
  \citenamefont {Arita}, \citenamefont {Namatame}, \citenamefont {Taniguchi},
  \citenamefont {Katayama}, \citenamefont {Nohara}, \citenamefont {Takagi},
  \citenamefont {Mizokawa},\ and\ \citenamefont {Ohta}}]{Seki2014}%
  \BibitemOpen
  \bibfield  {author} {\bibinfo {author} {\bibfnamefont {K.}~\bibnamefont
  {Seki}}, \bibinfo {author} {\bibfnamefont {Y.}~\bibnamefont {Wakisaka}},
  \bibinfo {author} {\bibfnamefont {T.}~\bibnamefont {Kaneko}}, \bibinfo
  {author} {\bibfnamefont {T.}~\bibnamefont {Toriyama}}, \bibinfo {author}
  {\bibfnamefont {T.}~\bibnamefont {Konishi}}, \bibinfo {author} {\bibfnamefont
  {T.}~\bibnamefont {Sudayama}}, \bibinfo {author} {\bibfnamefont {N.~L.}\
  \bibnamefont {Saini}}, \bibinfo {author} {\bibfnamefont {M.}~\bibnamefont
  {Arita}}, \bibinfo {author} {\bibfnamefont {H.}~\bibnamefont {Namatame}},
  \bibinfo {author} {\bibfnamefont {M.}~\bibnamefont {Taniguchi}}, \bibinfo
  {author} {\bibfnamefont {N.}~\bibnamefont {Katayama}}, \bibinfo {author}
  {\bibfnamefont {M.}~\bibnamefont {Nohara}}, \bibinfo {author} {\bibfnamefont
  {H.}~\bibnamefont {Takagi}}, \bibinfo {author} {\bibfnamefont
  {T.}~\bibnamefont {Mizokawa}}, \ and\ \bibinfo {author} {\bibfnamefont
  {Y.}~\bibnamefont {Ohta}},\ }\href@noop {} {\bibfield  {journal} {\bibinfo
  {journal} {Phys. Rev. B}\ }\textbf {\bibinfo {volume} {90}},\ \bibinfo
  {pages} {155116} (\bibinfo {year} {2014})}\BibitemShut {NoStop}%
\bibitem [{\citenamefont {Fukutani}\ \emph {et~al.}(2021)\citenamefont
  {Fukutani}, \citenamefont {Stania}, \citenamefont {Il~Kwon}, \citenamefont
  {Kim}, \citenamefont {Kong}, \citenamefont {Kim},\ and\ \citenamefont
  {Yeom}}]{Fukutani2021}%
  \BibitemOpen
  \bibfield  {author} {\bibinfo {author} {\bibfnamefont {K.}~\bibnamefont
  {Fukutani}}, \bibinfo {author} {\bibfnamefont {R.}~\bibnamefont {Stania}},
  \bibinfo {author} {\bibfnamefont {C.}~\bibnamefont {Il~Kwon}}, \bibinfo
  {author} {\bibfnamefont {J.~S.}\ \bibnamefont {Kim}}, \bibinfo {author}
  {\bibfnamefont {K.~J.}\ \bibnamefont {Kong}}, \bibinfo {author}
  {\bibfnamefont {J.}~\bibnamefont {Kim}}, \ and\ \bibinfo {author}
  {\bibfnamefont {H.~W.}\ \bibnamefont {Yeom}},\ }\href@noop {} {\bibfield
  {journal} {\bibinfo  {journal} {Nat. Phys.}\ }\textbf {\bibinfo {volume}
  {17}},\ \bibinfo {pages} {1024} (\bibinfo {year} {2021})}\BibitemShut
  {NoStop}%
\bibitem [{\citenamefont {Larkin}\ \emph {et~al.}(2017)\citenamefont {Larkin},
  \citenamefont {Yaresko}, \citenamefont {Pröpper}, \citenamefont {Kikoin},
  \citenamefont {Lu}, \citenamefont {Takayama}, \citenamefont {Mathis},
  \citenamefont {Rost}, \citenamefont {Takagi}, \citenamefont {Keimer},\ and\
  \citenamefont {Boris}}]{Larkin2017}%
  \BibitemOpen
  \bibfield  {author} {\bibinfo {author} {\bibfnamefont {T.~I.}\ \bibnamefont
  {Larkin}}, \bibinfo {author} {\bibfnamefont {A.~N.}\ \bibnamefont {Yaresko}},
  \bibinfo {author} {\bibfnamefont {D.}~\bibnamefont {Pröpper}}, \bibinfo
  {author} {\bibfnamefont {K.~A.}\ \bibnamefont {Kikoin}}, \bibinfo {author}
  {\bibfnamefont {Y.~F.}\ \bibnamefont {Lu}}, \bibinfo {author} {\bibfnamefont
  {T.}~\bibnamefont {Takayama}}, \bibinfo {author} {\bibfnamefont {Y.~L.}\
  \bibnamefont {Mathis}}, \bibinfo {author} {\bibfnamefont {A.~W.}\
  \bibnamefont {Rost}}, \bibinfo {author} {\bibfnamefont {H.}~\bibnamefont
  {Takagi}}, \bibinfo {author} {\bibfnamefont {B.}~\bibnamefont {Keimer}}, \
  and\ \bibinfo {author} {\bibfnamefont {A.~V.}\ \bibnamefont {Boris}},\
  }\href@noop {} {\bibfield  {journal} {\bibinfo  {journal} {Phys. Rev. B}\
  }\textbf {\bibinfo {volume} {95}},\ \bibinfo {pages} {195144} (\bibinfo
  {year} {2017})}\BibitemShut {NoStop}%
\bibitem [{\citenamefont {Larkin}\ \emph {et~al.}(2018)\citenamefont {Larkin},
  \citenamefont {Dawson}, \citenamefont {Höppner}, \citenamefont {Takayama},
  \citenamefont {Isobe}, \citenamefont {Mathis}, \citenamefont {Takagi},
  \citenamefont {Keimer},\ and\ \citenamefont {Boris}}]{Larkin2018}%
  \BibitemOpen
  \bibfield  {author} {\bibinfo {author} {\bibfnamefont {T.~I.}\ \bibnamefont
  {Larkin}}, \bibinfo {author} {\bibfnamefont {R.~D.}\ \bibnamefont {Dawson}},
  \bibinfo {author} {\bibfnamefont {M.}~\bibnamefont {Höppner}}, \bibinfo
  {author} {\bibfnamefont {T.}~\bibnamefont {Takayama}}, \bibinfo {author}
  {\bibfnamefont {M.}~\bibnamefont {Isobe}}, \bibinfo {author} {\bibfnamefont
  {Y.~L.}\ \bibnamefont {Mathis}}, \bibinfo {author} {\bibfnamefont
  {H.}~\bibnamefont {Takagi}}, \bibinfo {author} {\bibfnamefont
  {B.}~\bibnamefont {Keimer}}, \ and\ \bibinfo {author} {\bibfnamefont {A.~V.}\
  \bibnamefont {Boris}},\ }\href@noop {} {\bibfield  {journal} {\bibinfo
  {journal} {Phys. Rev. B}\ }\textbf {\bibinfo {volume} {98}},\ \bibinfo
  {pages} {125113} (\bibinfo {year} {2018})}\BibitemShut {NoStop}%
\bibitem [{\citenamefont {Sugimoto}\ \emph {et~al.}(2018)\citenamefont
  {Sugimoto}, \citenamefont {Nishimoto}, \citenamefont {Kaneko},\ and\
  \citenamefont {Ohta}}]{Sugimoto2018}%
  \BibitemOpen
  \bibfield  {author} {\bibinfo {author} {\bibfnamefont {K.}~\bibnamefont
  {Sugimoto}}, \bibinfo {author} {\bibfnamefont {S.}~\bibnamefont {Nishimoto}},
  \bibinfo {author} {\bibfnamefont {T.}~\bibnamefont {Kaneko}}, \ and\ \bibinfo
  {author} {\bibfnamefont {Y.}~\bibnamefont {Ohta}},\ }\href@noop {} {\bibfield
   {journal} {\bibinfo  {journal} {Phys. Rev. Lett.}\ }\textbf {\bibinfo
  {volume} {120}},\ \bibinfo {pages} {247602} (\bibinfo {year}
  {2018})}\BibitemShut {NoStop}%
\bibitem [{\citenamefont {Kaneko}\ \emph {et~al.}(2013)\citenamefont {Kaneko},
  \citenamefont {Toriyama}, \citenamefont {Konishi},\ and\ \citenamefont
  {Ohta}}]{Kaneko2013}%
  \BibitemOpen
  \bibfield  {author} {\bibinfo {author} {\bibfnamefont {T.}~\bibnamefont
  {Kaneko}}, \bibinfo {author} {\bibfnamefont {T.}~\bibnamefont {Toriyama}},
  \bibinfo {author} {\bibfnamefont {T.}~\bibnamefont {Konishi}}, \ and\
  \bibinfo {author} {\bibfnamefont {Y.}~\bibnamefont {Ohta}},\ }\href@noop {}
  {\bibfield  {journal} {\bibinfo  {journal} {Phys. Rev. B}\ }\textbf {\bibinfo
  {volume} {87}},\ \bibinfo {pages} {035121} (\bibinfo {year}
  {2013})}\BibitemShut {NoStop}%
\bibitem [{\citenamefont {Ejima}\ \emph {et~al.}(2014)\citenamefont {Ejima},
  \citenamefont {Kaneko}, \citenamefont {Ohta},\ and\ \citenamefont
  {Fehske}}]{Ejima2014}%
  \BibitemOpen
  \bibfield  {author} {\bibinfo {author} {\bibfnamefont {S.}~\bibnamefont
  {Ejima}}, \bibinfo {author} {\bibfnamefont {T.}~\bibnamefont {Kaneko}},
  \bibinfo {author} {\bibfnamefont {Y.}~\bibnamefont {Ohta}}, \ and\ \bibinfo
  {author} {\bibfnamefont {H.}~\bibnamefont {Fehske}},\ }\href@noop {}
  {\bibfield  {journal} {\bibinfo  {journal} {Phys. Rev. Lett.}\ }\textbf
  {\bibinfo {volume} {112}},\ \bibinfo {pages} {026401} (\bibinfo {year}
  {2014})}\BibitemShut {NoStop}%
\bibitem [{\citenamefont {Sugimoto}\ \emph {et~al.}(2016)\citenamefont
  {Sugimoto}, \citenamefont {Kaneko},\ and\ \citenamefont
  {Ohta}}]{Sugimoto2016}%
  \BibitemOpen
  \bibfield  {author} {\bibinfo {author} {\bibfnamefont {K.}~\bibnamefont
  {Sugimoto}}, \bibinfo {author} {\bibfnamefont {T.}~\bibnamefont {Kaneko}}, \
  and\ \bibinfo {author} {\bibfnamefont {Y.}~\bibnamefont {Ohta}},\ }\href@noop
  {} {\bibfield  {journal} {\bibinfo  {journal} {Phys. Rev. B}\ }\textbf
  {\bibinfo {volume} {93}},\ \bibinfo {pages} {041105} (\bibinfo {year}
  {2016})}\BibitemShut {NoStop}%
\bibitem [{\citenamefont {Yamada}\ \emph {et~al.}(2016)\citenamefont {Yamada},
  \citenamefont {Domon},\ and\ \citenamefont {Ōno}}]{Yamada2016}%
  \BibitemOpen
  \bibfield  {author} {\bibinfo {author} {\bibfnamefont {T.}~\bibnamefont
  {Yamada}}, \bibinfo {author} {\bibfnamefont {K.}~\bibnamefont {Domon}}, \
  and\ \bibinfo {author} {\bibfnamefont {Y.}~\bibnamefont {Ōno}},\ }\href@noop
  {} {\bibfield  {journal} {\bibinfo  {journal} {J. Phys. Soc. Japan}\ }\textbf
  {\bibinfo {volume} {85}},\ \bibinfo {pages} {053703} (\bibinfo {year}
  {2016})}\BibitemShut {NoStop}%
\bibitem [{\citenamefont {Kim}\ \emph {et~al.}(2021)\citenamefont {Kim},
  \citenamefont {Kim}, \citenamefont {Kim}, \citenamefont {Kwon}, \citenamefont
  {Kim},\ and\ \citenamefont {Kim}}]{Kim2021}%
  \BibitemOpen
  \bibfield  {author} {\bibinfo {author} {\bibfnamefont {K.}~\bibnamefont
  {Kim}}, \bibinfo {author} {\bibfnamefont {H.}~\bibnamefont {Kim}}, \bibinfo
  {author} {\bibfnamefont {J.}~\bibnamefont {Kim}}, \bibinfo {author}
  {\bibfnamefont {C.}~\bibnamefont {Kwon}}, \bibinfo {author} {\bibfnamefont
  {J.~S.}\ \bibnamefont {Kim}}, \ and\ \bibinfo {author} {\bibfnamefont
  {B.~J.}\ \bibnamefont {Kim}},\ }\href@noop {} {\bibfield  {journal} {\bibinfo
   {journal} {Nat. Commun.}\ }\textbf {\bibinfo {volume} {12}},\ \bibinfo
  {pages} {1969} (\bibinfo {year} {2021})}\BibitemShut {NoStop}%
\bibitem [{\citenamefont {Volkov}\ \emph {et~al.}(2021)\citenamefont {Volkov},
  \citenamefont {Ye}, \citenamefont {Lohani}, \citenamefont {Feldman},
  \citenamefont {Kanigel},\ and\ \citenamefont {Blumberg}}]{Volkov2021_npj}%
  \BibitemOpen
  \bibfield  {author} {\bibinfo {author} {\bibfnamefont {P.~A.}\ \bibnamefont
  {Volkov}}, \bibinfo {author} {\bibfnamefont {M.}~\bibnamefont {Ye}}, \bibinfo
  {author} {\bibfnamefont {H.}~\bibnamefont {Lohani}}, \bibinfo {author}
  {\bibfnamefont {I.}~\bibnamefont {Feldman}}, \bibinfo {author} {\bibfnamefont
  {A.}~\bibnamefont {Kanigel}}, \ and\ \bibinfo {author} {\bibfnamefont
  {G.}~\bibnamefont {Blumberg}},\ }\href@noop {} {\bibfield  {journal}
  {\bibinfo  {journal} {npj Quantum Mater.}\ }\textbf {\bibinfo {volume} {6}},\
  \bibinfo {pages} {52} (\bibinfo {year} {2021})}\BibitemShut {NoStop}%
\bibitem [{\citenamefont {Ye}\ \emph {et~al.}(2021)\citenamefont {Ye},
  \citenamefont {Volkov}, \citenamefont {Lohani}, \citenamefont {Feldman},
  \citenamefont {Kim}, \citenamefont {Kanigel},\ and\ \citenamefont
  {Blumberg}}]{Ye2021}%
  \BibitemOpen
  \bibfield  {author} {\bibinfo {author} {\bibfnamefont {M.}~\bibnamefont
  {Ye}}, \bibinfo {author} {\bibfnamefont {P.~A.}\ \bibnamefont {Volkov}},
  \bibinfo {author} {\bibfnamefont {H.}~\bibnamefont {Lohani}}, \bibinfo
  {author} {\bibfnamefont {I.}~\bibnamefont {Feldman}}, \bibinfo {author}
  {\bibfnamefont {M.}~\bibnamefont {Kim}}, \bibinfo {author} {\bibfnamefont
  {A.}~\bibnamefont {Kanigel}}, \ and\ \bibinfo {author} {\bibfnamefont
  {G.}~\bibnamefont {Blumberg}},\ }\href@noop {} {\bibfield  {journal}
  {\bibinfo  {journal} {Phys. Rev. B}\ }\textbf {\bibinfo {volume} {104}},\
  \bibinfo {pages} {045102} (\bibinfo {year} {2021})}\BibitemShut {NoStop}%
\bibitem [{\citenamefont {Nakano}\ \emph {et~al.}(2018)\citenamefont {Nakano},
  \citenamefont {Hasegawa}, \citenamefont {Tamura}, \citenamefont {Katayama},
  \citenamefont {Tsutsui},\ and\ \citenamefont {Sawa}}]{Nakano2018}%
  \BibitemOpen
  \bibfield  {author} {\bibinfo {author} {\bibfnamefont {A.}~\bibnamefont
  {Nakano}}, \bibinfo {author} {\bibfnamefont {T.}~\bibnamefont {Hasegawa}},
  \bibinfo {author} {\bibfnamefont {S.}~\bibnamefont {Tamura}}, \bibinfo
  {author} {\bibfnamefont {N.}~\bibnamefont {Katayama}}, \bibinfo {author}
  {\bibfnamefont {S.}~\bibnamefont {Tsutsui}}, \ and\ \bibinfo {author}
  {\bibfnamefont {H.}~\bibnamefont {Sawa}},\ }\href@noop {} {\bibfield
  {journal} {\bibinfo  {journal} {Phys. Rev. B}\ }\textbf {\bibinfo {volume}
  {98}},\ \bibinfo {pages} {045139} (\bibinfo {year} {2018})}\BibitemShut
  {NoStop}%
\bibitem [{\citenamefont {Mazza}\ \emph {et~al.}(2020)\citenamefont {Mazza},
  \citenamefont {Rosner}, \citenamefont {Windg\"{a}tter}, \citenamefont
  {Latini}, \citenamefont {Hubener}, \citenamefont {Millis}, \citenamefont
  {Rubio},\ and\ \citenamefont {Georges}}]{Mazza2020}%
  \BibitemOpen
  \bibfield  {author} {\bibinfo {author} {\bibfnamefont {G.}~\bibnamefont
  {Mazza}}, \bibinfo {author} {\bibfnamefont {M.}~\bibnamefont {Rosner}},
  \bibinfo {author} {\bibfnamefont {L.}~\bibnamefont {Windg\"{a}tter}},
  \bibinfo {author} {\bibfnamefont {S.}~\bibnamefont {Latini}}, \bibinfo
  {author} {\bibfnamefont {H.}~\bibnamefont {Hubener}}, \bibinfo {author}
  {\bibfnamefont {A.~J.}\ \bibnamefont {Millis}}, \bibinfo {author}
  {\bibfnamefont {A.}~\bibnamefont {Rubio}}, \ and\ \bibinfo {author}
  {\bibfnamefont {A.}~\bibnamefont {Georges}},\ }\href@noop {} {\bibfield
  {journal} {\bibinfo  {journal} {Phys. Rev. Lett.}\ }\textbf {\bibinfo
  {volume} {124}},\ \bibinfo {pages} {197601} (\bibinfo {year}
  {2020})}\BibitemShut {NoStop}%
\bibitem [{\citenamefont {Watson}\ \emph {et~al.}(2020)\citenamefont {Watson},
  \citenamefont {Markovi\'{c}}, \citenamefont {Morales}, \citenamefont
  {Le~F\`{e}vre}, \citenamefont {Merz}, \citenamefont {Haghighirad},\ and\
  \citenamefont {King}}]{Watson2020}%
  \BibitemOpen
  \bibfield  {author} {\bibinfo {author} {\bibfnamefont {M.~D.}\ \bibnamefont
  {Watson}}, \bibinfo {author} {\bibfnamefont {I.}~\bibnamefont
  {Markovi\'{c}}}, \bibinfo {author} {\bibfnamefont {E.~A.}\ \bibnamefont
  {Morales}}, \bibinfo {author} {\bibfnamefont {P.}~\bibnamefont
  {Le~F\`{e}vre}}, \bibinfo {author} {\bibfnamefont {M.}~\bibnamefont {Merz}},
  \bibinfo {author} {\bibfnamefont {A.~A.}\ \bibnamefont {Haghighirad}}, \ and\
  \bibinfo {author} {\bibfnamefont {P.~D.~C.}\ \bibnamefont {King}},\
  }\href@noop {} {\bibfield  {journal} {\bibinfo  {journal} {Phys. Rev. Res.}\
  }\textbf {\bibinfo {volume} {2}},\ \bibinfo {pages} {013236} (\bibinfo {year}
  {2020})}\BibitemShut {NoStop}%
\bibitem [{\citenamefont {Zenker}\ \emph {et~al.}(2014)\citenamefont {Zenker},
  \citenamefont {Fehske},\ and\ \citenamefont {Beck}}]{zenker_fate_2014}%
  \BibitemOpen
  \bibfield  {author} {\bibinfo {author} {\bibfnamefont {B.}~\bibnamefont
  {Zenker}}, \bibinfo {author} {\bibfnamefont {H.}~\bibnamefont {Fehske}}, \
  and\ \bibinfo {author} {\bibfnamefont {H.}~\bibnamefont {Beck}},\ }\href@noop
  {} {\bibfield  {journal} {\bibinfo  {journal} {Phys. Rev. B}\ }\textbf
  {\bibinfo {volume} {90}},\ \bibinfo {pages} {195118} (\bibinfo {year}
  {2014})}\BibitemShut {NoStop}%
\bibitem [{\citenamefont {Subedi}(2020)}]{Subedi2020}%
  \BibitemOpen
  \bibfield  {author} {\bibinfo {author} {\bibfnamefont {A.}~\bibnamefont
  {Subedi}},\ }\href@noop {} {\bibfield  {journal} {\bibinfo  {journal} {Phys.
  Rev. Mater.}\ }\textbf {\bibinfo {volume} {4}},\ \bibinfo {pages} {083601}
  (\bibinfo {year} {2020})}\BibitemShut {NoStop}%
\bibitem [{\citenamefont {Windg\"{a}tter}\ \emph {et~al.}(2021)\citenamefont
  {Windg\"{a}tter}, \citenamefont {R\"{o}sner}, \citenamefont {Mazza},
  \citenamefont {H\"{u}bener}, \citenamefont {Georges}, \citenamefont {Millis},
  \citenamefont {Latini},\ and\ \citenamefont {Rubio}}]{Windgatter2021}%
  \BibitemOpen
  \bibfield  {author} {\bibinfo {author} {\bibfnamefont {L.}~\bibnamefont
  {Windg\"{a}tter}}, \bibinfo {author} {\bibfnamefont {M.}~\bibnamefont
  {R\"{o}sner}}, \bibinfo {author} {\bibfnamefont {G.}~\bibnamefont {Mazza}},
  \bibinfo {author} {\bibfnamefont {H.}~\bibnamefont {H\"{u}bener}}, \bibinfo
  {author} {\bibfnamefont {A.}~\bibnamefont {Georges}}, \bibinfo {author}
  {\bibfnamefont {A.~J.}\ \bibnamefont {Millis}}, \bibinfo {author}
  {\bibfnamefont {S.}~\bibnamefont {Latini}}, \ and\ \bibinfo {author}
  {\bibfnamefont {A.}~\bibnamefont {Rubio}},\ }\href@noop {} {\bibfield
  {journal} {\bibinfo  {journal} {npj Comput. Mater.}\ }\textbf {\bibinfo
  {volume} {7}},\ \bibinfo {pages} {210} (\bibinfo {year} {2021})}\BibitemShut
  {NoStop}%
\bibitem [{\citenamefont {Hellmann}\ \emph {et~al.}(2012)\citenamefont
  {Hellmann}, \citenamefont {Rohwer}, \citenamefont {Kalläne}, \citenamefont
  {Hanff}, \citenamefont {Sohrt}, \citenamefont {Stange}, \citenamefont {Carr},
  \citenamefont {Murnane}, \citenamefont {Kapteyn}, \citenamefont {Kipp},
  \citenamefont {Bauer},\ and\ \citenamefont
  {Rossnagel}}]{hellmann_time-domain_2012}%
  \BibitemOpen
  \bibfield  {author} {\bibinfo {author} {\bibfnamefont {S.}~\bibnamefont
  {Hellmann}}, \bibinfo {author} {\bibfnamefont {T.}~\bibnamefont {Rohwer}},
  \bibinfo {author} {\bibfnamefont {M.}~\bibnamefont {Kalläne}}, \bibinfo
  {author} {\bibfnamefont {K.}~\bibnamefont {Hanff}}, \bibinfo {author}
  {\bibfnamefont {C.}~\bibnamefont {Sohrt}}, \bibinfo {author} {\bibfnamefont
  {A.}~\bibnamefont {Stange}}, \bibinfo {author} {\bibfnamefont
  {A.}~\bibnamefont {Carr}}, \bibinfo {author} {\bibfnamefont {M.~M.}\
  \bibnamefont {Murnane}}, \bibinfo {author} {\bibfnamefont {H.~C.}\
  \bibnamefont {Kapteyn}}, \bibinfo {author} {\bibfnamefont {L.}~\bibnamefont
  {Kipp}}, \bibinfo {author} {\bibfnamefont {M.}~\bibnamefont {Bauer}}, \ and\
  \bibinfo {author} {\bibfnamefont {K.}~\bibnamefont {Rossnagel}},\ }\href@noop
  {} {\bibfield  {journal} {\bibinfo  {journal} {Nat. Commun.}\ }\textbf
  {\bibinfo {volume} {3}},\ \bibinfo {pages} {1069} (\bibinfo {year}
  {2012})}\BibitemShut {NoStop}%
\bibitem [{\citenamefont {Giannetti}\ \emph {et~al.}(2016)\citenamefont
  {Giannetti}, \citenamefont {Capone}, \citenamefont {Fausti}, \citenamefont
  {Fabrizio}, \citenamefont {Parmigiani},\ and\ \citenamefont
  {Mihailovic}}]{Gianetti2016}%
  \BibitemOpen
  \bibfield  {author} {\bibinfo {author} {\bibfnamefont {C.}~\bibnamefont
  {Giannetti}}, \bibinfo {author} {\bibfnamefont {M.}~\bibnamefont {Capone}},
  \bibinfo {author} {\bibfnamefont {D.}~\bibnamefont {Fausti}}, \bibinfo
  {author} {\bibfnamefont {M.}~\bibnamefont {Fabrizio}}, \bibinfo {author}
  {\bibfnamefont {F.}~\bibnamefont {Parmigiani}}, \ and\ \bibinfo {author}
  {\bibfnamefont {D.}~\bibnamefont {Mihailovic}},\ }\href@noop {} {\bibfield
  {journal} {\bibinfo  {journal} {Advances in Physics}\ }\textbf {\bibinfo
  {volume} {65}},\ \bibinfo {pages} {58} (\bibinfo {year} {2016})}\BibitemShut
  {NoStop}%
\bibitem [{\citenamefont {Lloyd-Hughes}\ \emph {et~al.}(2021)\citenamefont
  {Lloyd-Hughes}, \citenamefont {Oppeneer}, \citenamefont {dos Santos},
  \citenamefont {Schleife}, \citenamefont {Meng}, \citenamefont {Sentef},
  \citenamefont {Ruggenthaler}, \citenamefont {Rubio}, \citenamefont {Radu},
  \citenamefont {Murnane}, \citenamefont {Shi}, \citenamefont {Kapteyn},
  \citenamefont {Stadtmüller}, \citenamefont {Dani}, \citenamefont
  {da~Jornada}, \citenamefont {Prinz}, \citenamefont {Aeschlimann},
  \citenamefont {Milot}, \citenamefont {Burdanova}, \citenamefont {Boland},
  \citenamefont {Cocker},\ and\ \citenamefont {Hegmann}}]{Lloyd_Hughes_2021}%
  \BibitemOpen
  \bibfield  {author} {\bibinfo {author} {\bibfnamefont {J.}~\bibnamefont
  {Lloyd-Hughes}}, \bibinfo {author} {\bibfnamefont {P.~M.}\ \bibnamefont
  {Oppeneer}}, \bibinfo {author} {\bibfnamefont {T.~P.}\ \bibnamefont {dos
  Santos}}, \bibinfo {author} {\bibfnamefont {A.}~\bibnamefont {Schleife}},
  \bibinfo {author} {\bibfnamefont {S.}~\bibnamefont {Meng}}, \bibinfo {author}
  {\bibfnamefont {M.~A.}\ \bibnamefont {Sentef}}, \bibinfo {author}
  {\bibfnamefont {M.}~\bibnamefont {Ruggenthaler}}, \bibinfo {author}
  {\bibfnamefont {A.}~\bibnamefont {Rubio}}, \bibinfo {author} {\bibfnamefont
  {I.}~\bibnamefont {Radu}}, \bibinfo {author} {\bibfnamefont {M.}~\bibnamefont
  {Murnane}}, \bibinfo {author} {\bibfnamefont {X.}~\bibnamefont {Shi}},
  \bibinfo {author} {\bibfnamefont {H.}~\bibnamefont {Kapteyn}}, \bibinfo
  {author} {\bibfnamefont {B.}~\bibnamefont {Stadtmüller}}, \bibinfo {author}
  {\bibfnamefont {K.~M.}\ \bibnamefont {Dani}}, \bibinfo {author}
  {\bibfnamefont {F.~H.}\ \bibnamefont {da~Jornada}}, \bibinfo {author}
  {\bibfnamefont {E.}~\bibnamefont {Prinz}}, \bibinfo {author} {\bibfnamefont
  {M.}~\bibnamefont {Aeschlimann}}, \bibinfo {author} {\bibfnamefont {R.~L.}\
  \bibnamefont {Milot}}, \bibinfo {author} {\bibfnamefont {M.}~\bibnamefont
  {Burdanova}}, \bibinfo {author} {\bibfnamefont {J.}~\bibnamefont {Boland}},
  \bibinfo {author} {\bibfnamefont {T.}~\bibnamefont {Cocker}}, \ and\ \bibinfo
  {author} {\bibfnamefont {F.}~\bibnamefont {Hegmann}},\ }\href@noop {}
  {\bibfield  {journal} {\bibinfo  {journal} {J. Phys. Condens. Matter}\
  }\textbf {\bibinfo {volume} {33}},\ \bibinfo {pages} {353001} (\bibinfo
  {year} {2021})}\BibitemShut {NoStop}%
\bibitem [{\citenamefont {Mor}\ \emph {et~al.}(2018)\citenamefont {Mor},
  \citenamefont {Herzog}, \citenamefont {Noack}, \citenamefont {Katayama},
  \citenamefont {Nohara}, \citenamefont {Takagi}, \citenamefont {Trunschke},
  \citenamefont {Mizokawa}, \citenamefont {Monney},\ and\ \citenamefont
  {St\"{a}hler}}]{Mor2018}%
  \BibitemOpen
  \bibfield  {author} {\bibinfo {author} {\bibfnamefont {S.}~\bibnamefont
  {Mor}}, \bibinfo {author} {\bibfnamefont {M.}~\bibnamefont {Herzog}},
  \bibinfo {author} {\bibfnamefont {J.}~\bibnamefont {Noack}}, \bibinfo
  {author} {\bibfnamefont {N.}~\bibnamefont {Katayama}}, \bibinfo {author}
  {\bibfnamefont {M.}~\bibnamefont {Nohara}}, \bibinfo {author} {\bibfnamefont
  {H.}~\bibnamefont {Takagi}}, \bibinfo {author} {\bibfnamefont
  {A.}~\bibnamefont {Trunschke}}, \bibinfo {author} {\bibfnamefont
  {T.}~\bibnamefont {Mizokawa}}, \bibinfo {author} {\bibfnamefont
  {C.}~\bibnamefont {Monney}}, \ and\ \bibinfo {author} {\bibfnamefont
  {J.}~\bibnamefont {St\"{a}hler}},\ }\href@noop {} {\bibfield  {journal}
  {\bibinfo  {journal} {Phys. Rev. B}\ }\textbf {\bibinfo {volume} {97}},\
  \bibinfo {pages} {115154} (\bibinfo {year} {2018})}\BibitemShut {NoStop}%
\bibitem [{\citenamefont {Werdehausen}\ \emph {et~al.}(2018)\citenamefont
  {Werdehausen}, \citenamefont {Takayama}, \citenamefont {Höppner},
  \citenamefont {Albrecht}, \citenamefont {Rost~Andreas}, \citenamefont {Lu},
  \citenamefont {Manske}, \citenamefont {Takagi},\ and\ \citenamefont
  {Kaiser}}]{Werdehausen2018}%
  \BibitemOpen
  \bibfield  {author} {\bibinfo {author} {\bibfnamefont {D.}~\bibnamefont
  {Werdehausen}}, \bibinfo {author} {\bibfnamefont {T.}~\bibnamefont
  {Takayama}}, \bibinfo {author} {\bibfnamefont {M.}~\bibnamefont {Höppner}},
  \bibinfo {author} {\bibfnamefont {G.}~\bibnamefont {Albrecht}}, \bibinfo
  {author} {\bibfnamefont {W.}~\bibnamefont {Rost~Andreas}}, \bibinfo {author}
  {\bibfnamefont {Y.}~\bibnamefont {Lu}}, \bibinfo {author} {\bibfnamefont
  {D.}~\bibnamefont {Manske}}, \bibinfo {author} {\bibfnamefont
  {H.}~\bibnamefont {Takagi}}, \ and\ \bibinfo {author} {\bibfnamefont
  {S.}~\bibnamefont {Kaiser}},\ }\href@noop {} {\bibfield  {journal} {\bibinfo
  {journal} {Sci. Adv.}\ }\textbf {\bibinfo {volume} {4}},\ \bibinfo {pages}
  {eaap8652} (\bibinfo {year} {2018})}\BibitemShut {NoStop}%
\bibitem [{\citenamefont {Bretscher}\ \emph {et~al.}(2021)\citenamefont
  {Bretscher}, \citenamefont {Andrich}, \citenamefont {Telang}, \citenamefont
  {Singh}, \citenamefont {Harnagea}, \citenamefont {Sood},\ and\ \citenamefont
  {Rao}}]{Bretscher2021}%
  \BibitemOpen
  \bibfield  {author} {\bibinfo {author} {\bibfnamefont {H.~M.}\ \bibnamefont
  {Bretscher}}, \bibinfo {author} {\bibfnamefont {P.}~\bibnamefont {Andrich}},
  \bibinfo {author} {\bibfnamefont {P.}~\bibnamefont {Telang}}, \bibinfo
  {author} {\bibfnamefont {A.}~\bibnamefont {Singh}}, \bibinfo {author}
  {\bibfnamefont {L.}~\bibnamefont {Harnagea}}, \bibinfo {author}
  {\bibfnamefont {A.~K.}\ \bibnamefont {Sood}}, \ and\ \bibinfo {author}
  {\bibfnamefont {A.}~\bibnamefont {Rao}},\ }\href@noop {} {\bibfield
  {journal} {\bibinfo  {journal} {Nat. Commun.}\ }\textbf {\bibinfo {volume}
  {12}},\ \bibinfo {pages} {1699} (\bibinfo {year} {2021})}\BibitemShut
  {NoStop}%
\bibitem [{\citenamefont {Liu}\ \emph {et~al.}(2021)\citenamefont {Liu},
  \citenamefont {Wu}, \citenamefont {Li}, \citenamefont {Shi}, \citenamefont
  {Wang}, \citenamefont {Zhang}, \citenamefont {Lin}, \citenamefont {Hu},
  \citenamefont {Tian}, \citenamefont {Li}, \citenamefont {Dong},\ and\
  \citenamefont {Wang}}]{Liu2021}%
  \BibitemOpen
  \bibfield  {author} {\bibinfo {author} {\bibfnamefont {Q.~M.}\ \bibnamefont
  {Liu}}, \bibinfo {author} {\bibfnamefont {D.}~\bibnamefont {Wu}}, \bibinfo
  {author} {\bibfnamefont {Z.~A.}\ \bibnamefont {Li}}, \bibinfo {author}
  {\bibfnamefont {L.~Y.}\ \bibnamefont {Shi}}, \bibinfo {author} {\bibfnamefont
  {Z.~X.}\ \bibnamefont {Wang}}, \bibinfo {author} {\bibfnamefont {S.~J.}\
  \bibnamefont {Zhang}}, \bibinfo {author} {\bibfnamefont {T.}~\bibnamefont
  {Lin}}, \bibinfo {author} {\bibfnamefont {T.~C.}\ \bibnamefont {Hu}},
  \bibinfo {author} {\bibfnamefont {H.~F.}\ \bibnamefont {Tian}}, \bibinfo
  {author} {\bibfnamefont {J.~Q.}\ \bibnamefont {Li}}, \bibinfo {author}
  {\bibfnamefont {T.}~\bibnamefont {Dong}}, \ and\ \bibinfo {author}
  {\bibfnamefont {N.~L.}\ \bibnamefont {Wang}},\ }\href@noop {} {\bibfield
  {journal} {\bibinfo  {journal} {Nat. Commun.}\ }\textbf {\bibinfo {volume}
  {12}},\ \bibinfo {pages} {2050} (\bibinfo {year} {2021})}\BibitemShut
  {NoStop}%
\bibitem [{\citenamefont {Miyamoto}\ \emph {et~al.}(2022)\citenamefont
  {Miyamoto}, \citenamefont {Mizui}, \citenamefont {Takamura}, \citenamefont
  {Hirata}, \citenamefont {Yamakawa}, \citenamefont {Morimoto}, \citenamefont
  {Terashige}, \citenamefont {Kida}, \citenamefont {Nakano}, \citenamefont
  {Sawa},\ and\ \citenamefont {Okamoto}}]{Miyamoto2022}%
  \BibitemOpen
  \bibfield  {author} {\bibinfo {author} {\bibfnamefont {T.}~\bibnamefont
  {Miyamoto}}, \bibinfo {author} {\bibfnamefont {M.}~\bibnamefont {Mizui}},
  \bibinfo {author} {\bibfnamefont {N.}~\bibnamefont {Takamura}}, \bibinfo
  {author} {\bibfnamefont {J.}~\bibnamefont {Hirata}}, \bibinfo {author}
  {\bibfnamefont {H.}~\bibnamefont {Yamakawa}}, \bibinfo {author}
  {\bibfnamefont {T.}~\bibnamefont {Morimoto}}, \bibinfo {author}
  {\bibfnamefont {T.}~\bibnamefont {Terashige}}, \bibinfo {author}
  {\bibfnamefont {N.}~\bibnamefont {Kida}}, \bibinfo {author} {\bibfnamefont
  {A.}~\bibnamefont {Nakano}}, \bibinfo {author} {\bibfnamefont
  {H.}~\bibnamefont {Sawa}}, \ and\ \bibinfo {author} {\bibfnamefont
  {H.}~\bibnamefont {Okamoto}},\ }\href@noop {} {\bibfield  {journal} {\bibinfo
   {journal} {J. Phys. Soc. Japan}\ }\textbf {\bibinfo {volume} {91}},\
  \bibinfo {pages} {023701} (\bibinfo {year} {2022})}\BibitemShut {NoStop}%
\bibitem [{\citenamefont {Mor}\ \emph {et~al.}(2017)\citenamefont {Mor},
  \citenamefont {Herzog}, \citenamefont {Golez}, \citenamefont {Werner},
  \citenamefont {Eckstein}, \citenamefont {Katayama}, \citenamefont {Nohara},
  \citenamefont {Takagi}, \citenamefont {Mizokawa}, \citenamefont {Monney},\
  and\ \citenamefont {Stahler}}]{Mor2017}%
  \BibitemOpen
  \bibfield  {author} {\bibinfo {author} {\bibfnamefont {S.}~\bibnamefont
  {Mor}}, \bibinfo {author} {\bibfnamefont {M.}~\bibnamefont {Herzog}},
  \bibinfo {author} {\bibfnamefont {D.}~\bibnamefont {Golez}}, \bibinfo
  {author} {\bibfnamefont {P.}~\bibnamefont {Werner}}, \bibinfo {author}
  {\bibfnamefont {M.}~\bibnamefont {Eckstein}}, \bibinfo {author}
  {\bibfnamefont {N.}~\bibnamefont {Katayama}}, \bibinfo {author}
  {\bibfnamefont {M.}~\bibnamefont {Nohara}}, \bibinfo {author} {\bibfnamefont
  {H.}~\bibnamefont {Takagi}}, \bibinfo {author} {\bibfnamefont
  {T.}~\bibnamefont {Mizokawa}}, \bibinfo {author} {\bibfnamefont
  {C.}~\bibnamefont {Monney}}, \ and\ \bibinfo {author} {\bibfnamefont
  {J.}~\bibnamefont {Stahler}},\ }\href@noop {} {\bibfield  {journal} {\bibinfo
   {journal} {Phys. Rev. Lett.}\ }\textbf {\bibinfo {volume} {119}},\ \bibinfo
  {pages} {086401} (\bibinfo {year} {2017})}\BibitemShut {NoStop}%
\bibitem [{\citenamefont {Okazaki}\ \emph {et~al.}(2018)\citenamefont
  {Okazaki}, \citenamefont {Ogawa}, \citenamefont {Suzuki}, \citenamefont
  {Yamamoto}, \citenamefont {Someya}, \citenamefont {Michimae}, \citenamefont
  {Watanabe}, \citenamefont {Lu}, \citenamefont {Nohara}, \citenamefont
  {Takagi}, \citenamefont {Katayama}, \citenamefont {Sawa}, \citenamefont
  {Fujisawa}, \citenamefont {Kanai}, \citenamefont {Ishii}, \citenamefont
  {Itatani}, \citenamefont {Mizokawa},\ and\ \citenamefont
  {Shin}}]{Okazaki2018}%
  \BibitemOpen
  \bibfield  {author} {\bibinfo {author} {\bibfnamefont {K.}~\bibnamefont
  {Okazaki}}, \bibinfo {author} {\bibfnamefont {Y.}~\bibnamefont {Ogawa}},
  \bibinfo {author} {\bibfnamefont {T.}~\bibnamefont {Suzuki}}, \bibinfo
  {author} {\bibfnamefont {T.}~\bibnamefont {Yamamoto}}, \bibinfo {author}
  {\bibfnamefont {T.}~\bibnamefont {Someya}}, \bibinfo {author} {\bibfnamefont
  {S.}~\bibnamefont {Michimae}}, \bibinfo {author} {\bibfnamefont
  {M.}~\bibnamefont {Watanabe}}, \bibinfo {author} {\bibfnamefont
  {Y.}~\bibnamefont {Lu}}, \bibinfo {author} {\bibfnamefont {M.}~\bibnamefont
  {Nohara}}, \bibinfo {author} {\bibfnamefont {H.}~\bibnamefont {Takagi}},
  \bibinfo {author} {\bibfnamefont {N.}~\bibnamefont {Katayama}}, \bibinfo
  {author} {\bibfnamefont {H.}~\bibnamefont {Sawa}}, \bibinfo {author}
  {\bibfnamefont {M.}~\bibnamefont {Fujisawa}}, \bibinfo {author}
  {\bibfnamefont {T.}~\bibnamefont {Kanai}}, \bibinfo {author} {\bibfnamefont
  {N.}~\bibnamefont {Ishii}}, \bibinfo {author} {\bibfnamefont
  {J.}~\bibnamefont {Itatani}}, \bibinfo {author} {\bibfnamefont
  {T.}~\bibnamefont {Mizokawa}}, \ and\ \bibinfo {author} {\bibfnamefont
  {S.}~\bibnamefont {Shin}},\ }\href@noop {} {\bibfield  {journal} {\bibinfo
  {journal} {Nat. Commun.}\ }\textbf {\bibinfo {volume} {9}},\ \bibinfo {pages}
  {4322} (\bibinfo {year} {2018})}\BibitemShut {NoStop}%
\bibitem [{\citenamefont {Tang}\ \emph {et~al.}(2020)\citenamefont {Tang},
  \citenamefont {Wang}, \citenamefont {Duan}, \citenamefont {Yang},
  \citenamefont {Huang}, \citenamefont {Guo}, \citenamefont {Qian},\ and\
  \citenamefont {Zhang}}]{Tang2020}%
  \BibitemOpen
  \bibfield  {author} {\bibinfo {author} {\bibfnamefont {T.}~\bibnamefont
  {Tang}}, \bibinfo {author} {\bibfnamefont {H.}~\bibnamefont {Wang}}, \bibinfo
  {author} {\bibfnamefont {S.}~\bibnamefont {Duan}}, \bibinfo {author}
  {\bibfnamefont {Y.}~\bibnamefont {Yang}}, \bibinfo {author} {\bibfnamefont
  {C.}~\bibnamefont {Huang}}, \bibinfo {author} {\bibfnamefont
  {Y.}~\bibnamefont {Guo}}, \bibinfo {author} {\bibfnamefont {D.}~\bibnamefont
  {Qian}}, \ and\ \bibinfo {author} {\bibfnamefont {W.}~\bibnamefont {Zhang}},\
  }\href@noop {} {\bibfield  {journal} {\bibinfo  {journal} {Phys. Rev. B}\
  }\textbf {\bibinfo {volume} {101}},\ \bibinfo {pages} {235148} (\bibinfo
  {year} {2020})}\BibitemShut {NoStop}%
\bibitem [{\citenamefont {Baldini}\ \emph {et~al.}(2020)\citenamefont
  {Baldini}, \citenamefont {Zong}, \citenamefont {Choi}, \citenamefont {Lee},
  \citenamefont {Michael}, \citenamefont {Windgaetter}, \citenamefont {Mazin},
  \citenamefont {Latini}, \citenamefont {Azoury}, \citenamefont {Lv} \emph
  {et~al.}}]{Baldini2020}%
  \BibitemOpen
  \bibfield  {author} {\bibinfo {author} {\bibfnamefont {E.}~\bibnamefont
  {Baldini}}, \bibinfo {author} {\bibfnamefont {A.}~\bibnamefont {Zong}},
  \bibinfo {author} {\bibfnamefont {D.}~\bibnamefont {Choi}}, \bibinfo {author}
  {\bibfnamefont {C.}~\bibnamefont {Lee}}, \bibinfo {author} {\bibfnamefont
  {M.~H.}\ \bibnamefont {Michael}}, \bibinfo {author} {\bibfnamefont
  {L.}~\bibnamefont {Windgaetter}}, \bibinfo {author} {\bibfnamefont {I.~I.}\
  \bibnamefont {Mazin}}, \bibinfo {author} {\bibfnamefont {S.}~\bibnamefont
  {Latini}}, \bibinfo {author} {\bibfnamefont {D.}~\bibnamefont {Azoury}},
  \bibinfo {author} {\bibfnamefont {B.}~\bibnamefont {Lv}},  \emph {et~al.},\
  }\href@noop {} {\bibfield  {journal} {\bibinfo  {journal} {arXiv preprint
  arXiv:2007.02909}\ } (\bibinfo {year} {2020})}\BibitemShut {NoStop}%
\bibitem [{\citenamefont {Suzuki}\ \emph {et~al.}(2021)\citenamefont {Suzuki},
  \citenamefont {Shinohara}, \citenamefont {Lu}, \citenamefont {Watanabe},
  \citenamefont {Xu}, \citenamefont {Ishikawa}, \citenamefont {Takagi},
  \citenamefont {Nohara}, \citenamefont {Katayama}, \citenamefont {Sawa},
  \citenamefont {Fujisawa}, \citenamefont {Kanai}, \citenamefont {Itatani},
  \citenamefont {Mizokawa}, \citenamefont {Shin},\ and\ \citenamefont
  {Okazaki}}]{Suzuki2021}%
  \BibitemOpen
  \bibfield  {author} {\bibinfo {author} {\bibfnamefont {T.}~\bibnamefont
  {Suzuki}}, \bibinfo {author} {\bibfnamefont {Y.}~\bibnamefont {Shinohara}},
  \bibinfo {author} {\bibfnamefont {Y.}~\bibnamefont {Lu}}, \bibinfo {author}
  {\bibfnamefont {M.}~\bibnamefont {Watanabe}}, \bibinfo {author}
  {\bibfnamefont {J.}~\bibnamefont {Xu}}, \bibinfo {author} {\bibfnamefont
  {K.~L.}\ \bibnamefont {Ishikawa}}, \bibinfo {author} {\bibfnamefont
  {H.}~\bibnamefont {Takagi}}, \bibinfo {author} {\bibfnamefont
  {M.}~\bibnamefont {Nohara}}, \bibinfo {author} {\bibfnamefont
  {N.}~\bibnamefont {Katayama}}, \bibinfo {author} {\bibfnamefont
  {H.}~\bibnamefont {Sawa}}, \bibinfo {author} {\bibfnamefont {M.}~\bibnamefont
  {Fujisawa}}, \bibinfo {author} {\bibfnamefont {T.}~\bibnamefont {Kanai}},
  \bibinfo {author} {\bibfnamefont {J.}~\bibnamefont {Itatani}}, \bibinfo
  {author} {\bibfnamefont {T.}~\bibnamefont {Mizokawa}}, \bibinfo {author}
  {\bibfnamefont {S.}~\bibnamefont {Shin}}, \ and\ \bibinfo {author}
  {\bibfnamefont {K.}~\bibnamefont {Okazaki}},\ }\href@noop {} {\bibfield
  {journal} {\bibinfo  {journal} {Phys. Rev. B}\ }\textbf {\bibinfo {volume}
  {103}},\ \bibinfo {pages} {L121105} (\bibinfo {year} {2021})}\BibitemShut
  {NoStop}%
\bibitem [{\citenamefont {Golez}\ \emph {et~al.}(2021)\citenamefont {Golez},
  \citenamefont {Dufresne}, \citenamefont {Kim}, \citenamefont {Boschini},
  \citenamefont {Chu}, \citenamefont {Murakami}, \citenamefont {Levy},
  \citenamefont {Mills}, \citenamefont {Zhdanovich},\ and\ \citenamefont
  {Isobe}}]{Golez2021}%
  \BibitemOpen
  \bibfield  {author} {\bibinfo {author} {\bibfnamefont {D.}~\bibnamefont
  {Golez}}, \bibinfo {author} {\bibfnamefont {S.~K.}\ \bibnamefont {Dufresne}},
  \bibinfo {author} {\bibfnamefont {M.-J.}\ \bibnamefont {Kim}}, \bibinfo
  {author} {\bibfnamefont {F.}~\bibnamefont {Boschini}}, \bibinfo {author}
  {\bibfnamefont {H.}~\bibnamefont {Chu}}, \bibinfo {author} {\bibfnamefont
  {Y.}~\bibnamefont {Murakami}}, \bibinfo {author} {\bibfnamefont
  {G.}~\bibnamefont {Levy}}, \bibinfo {author} {\bibfnamefont {A.~K.}\
  \bibnamefont {Mills}}, \bibinfo {author} {\bibfnamefont {S.}~\bibnamefont
  {Zhdanovich}}, \ and\ \bibinfo {author} {\bibfnamefont {M.}~\bibnamefont
  {Isobe}},\ }\href@noop {} {\bibfield  {journal} {\bibinfo  {journal} {arXiv
  preprint arXiv:2112.06298}\ } (\bibinfo {year} {2021})}\BibitemShut {NoStop}%
\bibitem [{\citenamefont {Saha}\ \emph {et~al.}(2021)\citenamefont {Saha},
  \citenamefont {Golež}, \citenamefont {De~Ninno}, \citenamefont {Mravlje},
  \citenamefont {Murakami}, \citenamefont {Ressel}, \citenamefont {Stupar},\
  and\ \citenamefont {Ribič}}]{Saha2021}%
  \BibitemOpen
  \bibfield  {author} {\bibinfo {author} {\bibfnamefont {T.}~\bibnamefont
  {Saha}}, \bibinfo {author} {\bibfnamefont {D.}~\bibnamefont {Golež}},
  \bibinfo {author} {\bibfnamefont {G.}~\bibnamefont {De~Ninno}}, \bibinfo
  {author} {\bibfnamefont {J.}~\bibnamefont {Mravlje}}, \bibinfo {author}
  {\bibfnamefont {Y.}~\bibnamefont {Murakami}}, \bibinfo {author}
  {\bibfnamefont {B.}~\bibnamefont {Ressel}}, \bibinfo {author} {\bibfnamefont
  {M.}~\bibnamefont {Stupar}}, \ and\ \bibinfo {author} {\bibfnamefont {P.~R.}\
  \bibnamefont {Ribič}},\ }\href@noop {} {\bibfield  {journal} {\bibinfo
  {journal} {Phys. Rev. B}\ }\textbf {\bibinfo {volume} {103}},\ \bibinfo
  {pages} {144304} (\bibinfo {year} {2021})}\BibitemShut {NoStop}%
\bibitem [{\citenamefont {Mor}\ \emph {et~al.}(2022)\citenamefont {Mor},
  \citenamefont {Herzog}, \citenamefont {Monney},\ and\ \citenamefont
  {St\"{a}hler}}]{Mor2022}%
  \BibitemOpen
  \bibfield  {author} {\bibinfo {author} {\bibfnamefont {S.}~\bibnamefont
  {Mor}}, \bibinfo {author} {\bibfnamefont {M.}~\bibnamefont {Herzog}},
  \bibinfo {author} {\bibfnamefont {C.}~\bibnamefont {Monney}}, \ and\ \bibinfo
  {author} {\bibfnamefont {J.}~\bibnamefont {St\"{a}hler}},\ }\href@noop {}
  {\bibfield  {journal} {\bibinfo  {journal} {Prog. Surf. Sci.}\ ,\ \bibinfo
  {pages} {100679}} (\bibinfo {year} {2022})}\BibitemShut {NoStop}%
\bibitem [{sm()}]{sm}%
  \BibitemOpen
  \href@noop {} {\bibinfo  {journal} {See Supplemental Material for the details
  of the experimental setup, data analysis, and the additional data}\
  }\BibitemShut {NoStop}%
\bibitem [{\citenamefont {Kim}\ \emph {et~al.}(2020)\citenamefont {Kim},
  \citenamefont {Schulz}, \citenamefont {Takayama}, \citenamefont {Isobe},
  \citenamefont {Takagi},\ and\ \citenamefont {Kaiser}}]{Kim2020}%
  \BibitemOpen
\bibfield  {journal} {  }\bibfield  {author} {\bibinfo {author} {\bibfnamefont
  {M.-J.}\ \bibnamefont {Kim}}, \bibinfo {author} {\bibfnamefont
  {A.}~\bibnamefont {Schulz}}, \bibinfo {author} {\bibfnamefont
  {T.}~\bibnamefont {Takayama}}, \bibinfo {author} {\bibfnamefont
  {M.}~\bibnamefont {Isobe}}, \bibinfo {author} {\bibfnamefont
  {H.}~\bibnamefont {Takagi}}, \ and\ \bibinfo {author} {\bibfnamefont
  {S.}~\bibnamefont {Kaiser}},\ }\href@noop {} {\bibfield  {journal} {\bibinfo
  {journal} {Phys. Rev. Res.}\ }\textbf {\bibinfo {volume} {2}},\ \bibinfo
  {pages} {042039(R)} (\bibinfo {year} {2020})}\BibitemShut {NoStop}%
\bibitem [{\citenamefont {Hayes}\ and\ \citenamefont {Loudon}()}]{Hayes2012}%
  \BibitemOpen
  \bibfield  {author} {\bibinfo {author} {\bibfnamefont {W.}~\bibnamefont
  {Hayes}}\ and\ \bibinfo {author} {\bibfnamefont {R.}~\bibnamefont {Loudon}},\
  }\href@noop {} {\emph {\bibinfo {title} {Scattering of Light by Crystals}}},\
  Dover Books on Physics\BibitemShut {NoStop}%
\bibitem [{\citenamefont {Yu}\ and\ \citenamefont {Cardona}(2001)}]{Yu}%
  \BibitemOpen
  \bibfield  {author} {\bibinfo {author} {\bibfnamefont {P.}~\bibnamefont
  {Yu}}\ and\ \bibinfo {author} {\bibfnamefont {M.}~\bibnamefont {Cardona}},\
  }\href@noop {} {\emph {\bibinfo {title} {Fundamentals of Semiconductors}}},\
  Advanced Text in Physics\ (\bibinfo  {publisher} {Springer},\ \bibinfo {year}
  {2001})\BibitemShut {NoStop}%
\bibitem [{\citenamefont {Fox}(2010)}]{Fox2010}%
  \BibitemOpen
  \bibfield  {author} {\bibinfo {author} {\bibfnamefont {M.}~\bibnamefont
  {Fox}},\ }\href@noop {} {\emph {\bibinfo {title} {Optical Properties of
  Solids}}},\ Oxford Master Series in Physics\ (\bibinfo  {publisher} {OUP
  Oxford},\ \bibinfo {year} {2010})\BibitemShut {NoStop}%
\end{thebibliography}%


\begin{thebibliography}{13}%
\makeatletter
\providecommand \@ifxundefined [1]{%
 \@ifx{#1\undefined}
}%
\providecommand \@ifnum [1]{%
 \ifnum #1\expandafter \@firstoftwo
 \else \expandafter \@secondoftwo
 \fi
}%
\providecommand \@ifx [1]{%
 \ifx #1\expandafter \@firstoftwo
 \else \expandafter \@secondoftwo
 \fi
}%
\providecommand \natexlab [1]{#1}%
\providecommand \enquote  [1]{``#1''}%
\providecommand \bibnamefont  [1]{#1}%
\providecommand \bibfnamefont [1]{#1}%
\providecommand \citenamefont [1]{#1}%
\providecommand \href@noop [0]{\@secondoftwo}%
\providecommand \href [0]{\begingroup \@sanitize@url \@href}%
\providecommand \@href[1]{\@@startlink{#1}\@@href}%
\providecommand \@@href[1]{\endgroup#1\@@endlink}%
\providecommand \@sanitize@url [0]{\catcode `\\12\catcode `\$12\catcode
  `\&12\catcode `\#12\catcode `\^12\catcode `\_12\catcode `\%12\relax}%
\providecommand \@@startlink[1]{}%
\providecommand \@@endlink[0]{}%
\providecommand \url  [0]{\begingroup\@sanitize@url \@url }%
\providecommand \@url [1]{\endgroup\@href {#1}{\urlprefix }}%
\providecommand \urlprefix  [0]{URL }%
\providecommand \Eprint [0]{\href }%
\providecommand \doibase [0]{http://dx.doi.org/}%
\providecommand \selectlanguage [0]{\@gobble}%
\providecommand \bibinfo  [0]{\@secondoftwo}%
\providecommand \bibfield  [0]{\@secondoftwo}%
\providecommand \translation [1]{[#1]}%
\providecommand \BibitemOpen [0]{}%
\providecommand \bibitemStop [0]{}%
\providecommand \bibitemNoStop [0]{.\EOS\space}%
\providecommand \EOS [0]{\spacefactor3000\relax}%
\providecommand \BibitemShut  [1]{\csname bibitem#1\endcsname}%
\let\auto@bib@innerbib\@empty
\bibitem [{\citenamefont {Sherldrick}(2008)}]{sherldrick}%
  \BibitemOpen
  \bibfield  {author} {\bibinfo {author} {\bibfnamefont {G.}~\bibnamefont
  {Sherldrick}},\ }\href@noop {} {\bibfield  {journal} {\bibinfo  {journal}
  {Acta Crystallogr., Sect. A: Found. Crystallogr.}\ }\textbf {\bibinfo
  {volume} {64}},\ \bibinfo {pages} {112} (\bibinfo {year} {2008})}\BibitemShut
  {NoStop}%
\bibitem [{\citenamefont {Petricek~V.}\ and\ \citenamefont
  {L.}(2014)}]{petricek}%
  \BibitemOpen
  \bibfield  {author} {\bibinfo {author} {\bibfnamefont {D.~M.}\ \bibnamefont
  {Petricek~V.}}\ and\ \bibinfo {author} {\bibfnamefont {P.}~\bibnamefont
  {L.}},\ }\href@noop {} {\bibfield  {journal} {\bibinfo  {journal} {Z.
  Kristallogr., Cryst. Mater.}\ }\textbf {\bibinfo {volume} {229}},\ \bibinfo
  {pages} {345} (\bibinfo {year} {2014})}\BibitemShut {NoStop}%
\bibitem [{\citenamefont {Nakano}\ \emph {et~al.}(2018)\citenamefont {Nakano},
  \citenamefont {Hasegawa}, \citenamefont {Tamura}, \citenamefont {Katayama},
  \citenamefont {Tsutsui},\ and\ \citenamefont {Sawa}}]{Nakano2018}%
  \BibitemOpen
  \bibfield  {author} {\bibinfo {author} {\bibfnamefont {A.}~\bibnamefont
  {Nakano}}, \bibinfo {author} {\bibfnamefont {T.}~\bibnamefont {Hasegawa}},
  \bibinfo {author} {\bibfnamefont {S.}~\bibnamefont {Tamura}}, \bibinfo
  {author} {\bibfnamefont {N.}~\bibnamefont {Katayama}}, \bibinfo {author}
  {\bibfnamefont {S.}~\bibnamefont {Tsutsui}}, \ and\ \bibinfo {author}
  {\bibfnamefont {H.}~\bibnamefont {Sawa}},\ }\href@noop {} {\bibfield
  {journal} {\bibinfo  {journal} {Phys. Rev. B}\ }\textbf {\bibinfo {volume}
  {98}},\ \bibinfo {pages} {045139} (\bibinfo {year} {2018})}\BibitemShut
  {NoStop}%
\bibitem [{\citenamefont {Hayes}\ and\ \citenamefont {Loudon}()}]{Hayes2012}%
  \BibitemOpen
  \bibfield  {author} {\bibinfo {author} {\bibfnamefont {W.}~\bibnamefont
  {Hayes}}\ and\ \bibinfo {author} {\bibfnamefont {R.}~\bibnamefont {Loudon}},\
  }\href@noop {} {\emph {\bibinfo {title} {Scattering of Light by Crystals}}},\
  Dover Books on Physics\BibitemShut {NoStop}%
\bibitem [{\citenamefont {Kim}\ \emph {et~al.}(2021)\citenamefont {Kim},
  \citenamefont {Kim}, \citenamefont {Kim}, \citenamefont {Kwon}, \citenamefont
  {Kim},\ and\ \citenamefont {Kim}}]{Kim2021}%
  \BibitemOpen
  \bibfield  {author} {\bibinfo {author} {\bibfnamefont {K.}~\bibnamefont
  {Kim}}, \bibinfo {author} {\bibfnamefont {H.}~\bibnamefont {Kim}}, \bibinfo
  {author} {\bibfnamefont {J.}~\bibnamefont {Kim}}, \bibinfo {author}
  {\bibfnamefont {C.}~\bibnamefont {Kwon}}, \bibinfo {author} {\bibfnamefont
  {J.~S.}\ \bibnamefont {Kim}}, \ and\ \bibinfo {author} {\bibfnamefont
  {B.~J.}\ \bibnamefont {Kim}},\ }\href@noop {} {\bibfield  {journal} {\bibinfo
   {journal} {Nat. Commun.}\ }\textbf {\bibinfo {volume} {12}},\ \bibinfo
  {pages} {1969} (\bibinfo {year} {2021})}\BibitemShut {NoStop}%
\bibitem [{\citenamefont {Volkov}\ \emph {et~al.}(2021)\citenamefont {Volkov},
  \citenamefont {Ye}, \citenamefont {Lohani}, \citenamefont {Feldman},
  \citenamefont {Kanigel},\ and\ \citenamefont {Blumberg}}]{Volkov2021_npj}%
  \BibitemOpen
  \bibfield  {author} {\bibinfo {author} {\bibfnamefont {P.~A.}\ \bibnamefont
  {Volkov}}, \bibinfo {author} {\bibfnamefont {M.}~\bibnamefont {Ye}}, \bibinfo
  {author} {\bibfnamefont {H.}~\bibnamefont {Lohani}}, \bibinfo {author}
  {\bibfnamefont {I.}~\bibnamefont {Feldman}}, \bibinfo {author} {\bibfnamefont
  {A.}~\bibnamefont {Kanigel}}, \ and\ \bibinfo {author} {\bibfnamefont
  {G.}~\bibnamefont {Blumberg}},\ }\href@noop {} {\bibfield  {journal}
  {\bibinfo  {journal} {npj Quantum Mater.}\ }\textbf {\bibinfo {volume} {6}},\
  \bibinfo {pages} {52} (\bibinfo {year} {2021})}\BibitemShut {NoStop}%
\bibitem [{\citenamefont {Ye}\ \emph {et~al.}(2021)\citenamefont {Ye},
  \citenamefont {Volkov}, \citenamefont {Lohani}, \citenamefont {Feldman},
  \citenamefont {Kim}, \citenamefont {Kanigel},\ and\ \citenamefont
  {Blumberg}}]{Ye2021}%
  \BibitemOpen
  \bibfield  {author} {\bibinfo {author} {\bibfnamefont {M.}~\bibnamefont
  {Ye}}, \bibinfo {author} {\bibfnamefont {P.~A.}\ \bibnamefont {Volkov}},
  \bibinfo {author} {\bibfnamefont {H.}~\bibnamefont {Lohani}}, \bibinfo
  {author} {\bibfnamefont {I.}~\bibnamefont {Feldman}}, \bibinfo {author}
  {\bibfnamefont {M.}~\bibnamefont {Kim}}, \bibinfo {author} {\bibfnamefont
  {A.}~\bibnamefont {Kanigel}}, \ and\ \bibinfo {author} {\bibfnamefont
  {G.}~\bibnamefont {Blumberg}},\ }\href@noop {} {\bibfield  {journal}
  {\bibinfo  {journal} {Phys. Rev. B}\ }\textbf {\bibinfo {volume} {104}},\
  \bibinfo {pages} {045102} (\bibinfo {year} {2021})}\BibitemShut {NoStop}%
\bibitem [{\citenamefont {Fox}(2010)}]{Fox2010}%
  \BibitemOpen
  \bibfield  {author} {\bibinfo {author} {\bibfnamefont {M.}~\bibnamefont
  {Fox}},\ }\href@noop {} {\emph {\bibinfo {title} {Optical Properties of
  Solids}}},\ Oxford Master Series in Physics\ (\bibinfo  {publisher} {OUP
  Oxford},\ \bibinfo {year} {2010})\BibitemShut {NoStop}%
\bibitem [{\citenamefont {Lu}\ \emph {et~al.}(2017)\citenamefont {Lu},
  \citenamefont {Kono}, \citenamefont {Larkin}, \citenamefont {Rost},
  \citenamefont {Takayama}, \citenamefont {Boris}, \citenamefont {Keimer},\
  and\ \citenamefont {Takagi}}]{Lu2017}%
  \BibitemOpen
  \bibfield  {author} {\bibinfo {author} {\bibfnamefont {Y.~F.}\ \bibnamefont
  {Lu}}, \bibinfo {author} {\bibfnamefont {H.}~\bibnamefont {Kono}}, \bibinfo
  {author} {\bibfnamefont {T.~I.}\ \bibnamefont {Larkin}}, \bibinfo {author}
  {\bibfnamefont {A.~W.}\ \bibnamefont {Rost}}, \bibinfo {author}
  {\bibfnamefont {T.}~\bibnamefont {Takayama}}, \bibinfo {author}
  {\bibfnamefont {A.~V.}\ \bibnamefont {Boris}}, \bibinfo {author}
  {\bibfnamefont {B.}~\bibnamefont {Keimer}}, \ and\ \bibinfo {author}
  {\bibfnamefont {H.}~\bibnamefont {Takagi}},\ }\href@noop {} {\bibfield
  {journal} {\bibinfo  {journal} {Nat. Commun.}\ }\textbf {\bibinfo {volume}
  {8}},\ \bibinfo {pages} {14408} (\bibinfo {year} {2017})}\BibitemShut
  {NoStop}%
\bibitem [{\citenamefont {Larkin}\ \emph {et~al.}(2017)\citenamefont {Larkin},
  \citenamefont {Yaresko}, \citenamefont {Pröpper}, \citenamefont {Kikoin},
  \citenamefont {Lu}, \citenamefont {Takayama}, \citenamefont {Mathis},
  \citenamefont {Rost}, \citenamefont {Takagi}, \citenamefont {Keimer},\ and\
  \citenamefont {Boris}}]{Larkin2017}%
  \BibitemOpen
  \bibfield  {author} {\bibinfo {author} {\bibfnamefont {T.~I.}\ \bibnamefont
  {Larkin}}, \bibinfo {author} {\bibfnamefont {A.~N.}\ \bibnamefont {Yaresko}},
  \bibinfo {author} {\bibfnamefont {D.}~\bibnamefont {Pröpper}}, \bibinfo
  {author} {\bibfnamefont {K.~A.}\ \bibnamefont {Kikoin}}, \bibinfo {author}
  {\bibfnamefont {Y.~F.}\ \bibnamefont {Lu}}, \bibinfo {author} {\bibfnamefont
  {T.}~\bibnamefont {Takayama}}, \bibinfo {author} {\bibfnamefont {Y.~L.}\
  \bibnamefont {Mathis}}, \bibinfo {author} {\bibfnamefont {A.~W.}\
  \bibnamefont {Rost}}, \bibinfo {author} {\bibfnamefont {H.}~\bibnamefont
  {Takagi}}, \bibinfo {author} {\bibfnamefont {B.}~\bibnamefont {Keimer}}, \
  and\ \bibinfo {author} {\bibfnamefont {A.~V.}\ \bibnamefont {Boris}},\
  }\href@noop {} {\bibfield  {journal} {\bibinfo  {journal} {Phys. Rev. B}\
  }\textbf {\bibinfo {volume} {95}},\ \bibinfo {pages} {195144} (\bibinfo
  {year} {2017})}\BibitemShut {NoStop}%
\bibitem [{\citenamefont {Larkin}\ \emph {et~al.}(2018)\citenamefont {Larkin},
  \citenamefont {Dawson}, \citenamefont {Höppner}, \citenamefont {Takayama},
  \citenamefont {Isobe}, \citenamefont {Mathis}, \citenamefont {Takagi},
  \citenamefont {Keimer},\ and\ \citenamefont {Boris}}]{Larkin2018}%
  \BibitemOpen
  \bibfield  {author} {\bibinfo {author} {\bibfnamefont {T.~I.}\ \bibnamefont
  {Larkin}}, \bibinfo {author} {\bibfnamefont {R.~D.}\ \bibnamefont {Dawson}},
  \bibinfo {author} {\bibfnamefont {M.}~\bibnamefont {Höppner}}, \bibinfo
  {author} {\bibfnamefont {T.}~\bibnamefont {Takayama}}, \bibinfo {author}
  {\bibfnamefont {M.}~\bibnamefont {Isobe}}, \bibinfo {author} {\bibfnamefont
  {Y.~L.}\ \bibnamefont {Mathis}}, \bibinfo {author} {\bibfnamefont
  {H.}~\bibnamefont {Takagi}}, \bibinfo {author} {\bibfnamefont
  {B.}~\bibnamefont {Keimer}}, \ and\ \bibinfo {author} {\bibfnamefont {A.~V.}\
  \bibnamefont {Boris}},\ }\href@noop {} {\bibfield  {journal} {\bibinfo
  {journal} {Phys. Rev. B}\ }\textbf {\bibinfo {volume} {98}},\ \bibinfo
  {pages} {125113} (\bibinfo {year} {2018})}\BibitemShut {NoStop}%
\bibitem [{\citenamefont {Sugimoto}\ \emph {et~al.}(2018)\citenamefont
  {Sugimoto}, \citenamefont {Nishimoto}, \citenamefont {Kaneko},\ and\
  \citenamefont {Ohta}}]{Sugimoto2018}%
  \BibitemOpen
  \bibfield  {author} {\bibinfo {author} {\bibfnamefont {K.}~\bibnamefont
  {Sugimoto}}, \bibinfo {author} {\bibfnamefont {S.}~\bibnamefont {Nishimoto}},
  \bibinfo {author} {\bibfnamefont {T.}~\bibnamefont {Kaneko}}, \ and\ \bibinfo
  {author} {\bibfnamefont {Y.}~\bibnamefont {Ohta}},\ }\href@noop {} {\bibfield
   {journal} {\bibinfo  {journal} {Phys. Rev. Lett.}\ }\textbf {\bibinfo
  {volume} {120}},\ \bibinfo {pages} {247602} (\bibinfo {year}
  {2018})}\BibitemShut {NoStop}%
\bibitem [{\citenamefont {Suzuki}\ \emph {et~al.}(2021)\citenamefont {Suzuki},
  \citenamefont {Shinohara}, \citenamefont {Lu}, \citenamefont {Watanabe},
  \citenamefont {Xu}, \citenamefont {Ishikawa}, \citenamefont {Takagi},
  \citenamefont {Nohara}, \citenamefont {Katayama}, \citenamefont {Sawa},
  \citenamefont {Fujisawa}, \citenamefont {Kanai}, \citenamefont {Itatani},
  \citenamefont {Mizokawa}, \citenamefont {Shin},\ and\ \citenamefont
  {Okazaki}}]{Suzuki2021}%
  \BibitemOpen
  \bibfield  {author} {\bibinfo {author} {\bibfnamefont {T.}~\bibnamefont
  {Suzuki}}, \bibinfo {author} {\bibfnamefont {Y.}~\bibnamefont {Shinohara}},
  \bibinfo {author} {\bibfnamefont {Y.}~\bibnamefont {Lu}}, \bibinfo {author}
  {\bibfnamefont {M.}~\bibnamefont {Watanabe}}, \bibinfo {author}
  {\bibfnamefont {J.}~\bibnamefont {Xu}}, \bibinfo {author} {\bibfnamefont
  {K.~L.}\ \bibnamefont {Ishikawa}}, \bibinfo {author} {\bibfnamefont
  {H.}~\bibnamefont {Takagi}}, \bibinfo {author} {\bibfnamefont
  {M.}~\bibnamefont {Nohara}}, \bibinfo {author} {\bibfnamefont
  {N.}~\bibnamefont {Katayama}}, \bibinfo {author} {\bibfnamefont
  {H.}~\bibnamefont {Sawa}}, \bibinfo {author} {\bibfnamefont {M.}~\bibnamefont
  {Fujisawa}}, \bibinfo {author} {\bibfnamefont {T.}~\bibnamefont {Kanai}},
  \bibinfo {author} {\bibfnamefont {J.}~\bibnamefont {Itatani}}, \bibinfo
  {author} {\bibfnamefont {T.}~\bibnamefont {Mizokawa}}, \bibinfo {author}
  {\bibfnamefont {S.}~\bibnamefont {Shin}}, \ and\ \bibinfo {author}
  {\bibfnamefont {K.}~\bibnamefont {Okazaki}},\ }\href@noop {} {\bibfield
  {journal} {\bibinfo  {journal} {Phys. Rev. B}\ }\textbf {\bibinfo {volume}
  {103}},\ \bibinfo {pages} {L121105} (\bibinfo {year} {2021})}\BibitemShut
  {NoStop}%
\end{thebibliography}%
	\bibliographystyle{apsrev4-1}
\end{document}


	\title{Disentangling lattice and electronic instabilities in the excitonic insulator candidate \tns by nonequilibrium spectroscopy\\
		Supplemental material}
	\author{Kota~Katsumi}
	\affiliation{Université Paris Cité, Matériaux et Phénomènes Quantiques UMR CNRQ 7162, Bâtiment Condorcet, 75205 Paris Cedex 13, France}
	\author{Alexandr~Alekhin}
	\affiliation{Université Paris Cité, Matériaux et Phénomènes Quantiques UMR CNRQ 7162, Bâtiment Condorcet, 75205 Paris Cedex 13, France}
	\author{Sofia-Michaela~Souliou}
	\affiliation{Institute for Quantum Materials and Technologies, Karlsruhe Institute of Technology, 76021 Karlsruhe, Germany}
	\author{Michael Merz}
	\affiliation{Institute for Quantum Materials and Technologies, Karlsruhe Institute of Technology, 76021 Karlsruhe, Germany}
	\affiliation{Karlsruhe Nano Micro Facility (KNMFi), Karlsruhe Institute of Technology, 76344 Eggenstein-Leopoldshafen, Germany}
	\author{Amir-Abbas~Haghighirad}
	\affiliation{Institute for Quantum Materials and Technologies, Karlsruhe Institute of Technology, 76021 Karlsruhe, Germany}
	\author{Matthieu~Le~Tacon}
	\affiliation{Institute for Quantum Materials and Technologies, Karlsruhe Institute of Technology, 76021 Karlsruhe, Germany}
	\author{Sarah~Houver}
	\affiliation{Université Paris Cité, Matériaux et Phénomènes Quantiques UMR CNRQ 7162, Bâtiment Condorcet, 75205 Paris Cedex 13, France}
	\author{Maximilien~Cazayous}
	\affiliation{Université Paris Cité, Matériaux et Phénomènes Quantiques UMR CNRQ 7162, Bâtiment Condorcet, 75205 Paris Cedex 13, France}
	\author{Alain~Sacuto}
	\affiliation{Université Paris Cité, Matériaux et Phénomènes Quantiques UMR CNRQ 7162, Bâtiment Condorcet, 75205 Paris Cedex 13, France}
	\author{Yann~Gallais}
	\affiliation{Université Paris Cité, Matériaux et Phénomènes Quantiques UMR CNRQ 7162, Bâtiment Condorcet, 75205 Paris Cedex 13, France}
	
	\maketitle
	\section{Methods}
	\subsection{Sample fabrication and characterization}
	High-quality single crystals of \tns (TNS) were grown by chemical vapor transport (CVT). Starting material in powder form was prepared utilizing a solid-state reaction. High-purity (at least 99.99\%) starting materials of Ta:Ni:Se with a ratio of 2:1:5 were mixed, weighed and subsequently reacted in a sealed fused silica ampule at 650~$^{\circ}$C for 7~days. The obtained product was ground in an agate mortar, mixed with iodine as a transport agent, and sealed in an evacuated quartz ampule. The quartz tube was placed in a horizontal two-zone tube furnace. The temperatures of the zones were kept at 950~$^{\circ}$C (i.e., for the source materials) and 750~$^{\circ}$C (for the growth region), respectively. The duration of the transport was 14~days. Needle-shaped single-crystals of typical dimensions 0.1~mm~$\times$~1~mm~$\times$~10~mm were formed in the cooler zone of the ampule. The composition of the crystals was confirmed by energy dispersive X-ray spectroscopy (EDS) device COXEM EM-30plus equipped with an Oxford detector.
	
	\subsection{X-ray diffraction measurements}
	A temperature-dependent X-ray diffraction (XRD) series on TNS single crystals from the same batch was conducted on a STOE imaging plate diffraction system (IPDS-2T) using Mo $K_{\alpha}$ radiation, and the transition temperature into the excitonic state was determined to $T_c{\rm(XRD)} = 326(1)$ K.\@ All accessible symmetry-equivalent reflections were measured up to a maximum angle of $2 \Theta =65^{\circ}$. The data were corrected for Lorentz, polarization, extinction, and absorption effects. Using SHELXL \cite{sherldrick} and JANA2006 \cite{petricek}, all averaged symmetry-independent reflections with $I > 2 \sigma$ (intensity of the reflections $I$ larger than the standard deviation of the counting statistics $\sigma$) have been included for the refinements in their corresponding space groups. Representative refinements for data collected at 365 K and 125 K are depicted in Table \ref{tab:table1}. For all temperatures, the refinement converged quite well and shows excellent reliability factors: see Goodness of Fit (GOF), and residues $R_1$,\@ and $wR_2$ in Table \ref{tab:table1}).\@
	
	\begin{table}[t]
		\caption{\label{tab:table1} Crystallographic data for TNS at 365~K and at 125~K determined from single-crystal X-ray diffraction. The high-temperature structure was refined in the orthorhombic space group (SG) $Cmcm$ and the low-temperature structure was refined in the monoclinic SG $C2/c$.\@ $U_{\rm equiv}$ denotes the equivalent atomic displacement parameters. The Wyckoff positions (WP) are given for the high- (HT) and low-temperature (LT) structures. TW represents the degree of twinning in the monoclinic phase. Note that the $\beta$ angle value reported here is lower than the value reported in earlier measurements \cite{Nakano2018} due to averaging over two twinned domains in the monoclinic phase.}
		\begin{ruledtabular}
			\begin{tabular}[t]{lllcc}
				& & Ta$_{2}$NiSe$_{5}$ & 365 K & 125 K \\
				\hline
				SG		& WP (HT/LT) &  & $Cmcm$ & $C2/c$ \\
				& & $a$ (\r{A}) & 3.4972(3) & 3.4874(7) \\
				& & $b$ (\r{A}) & 12.8487(15) & 12.8046(22) \\
				& & $c$ (\r{A}) & 15.6503(12) & 15.6336(19) \\
				& & $\beta$ $(^{\circ})$ & 90 & 90.123(10)\\
				Ta		& $8f$ & $x$ & 0 & -0.01136(4) \\
				& & $y$ & 0.22123(4) & 0.22142(5) \\
				& & $z$ & 0.11026(2) & 0.11064(4) \\
				& & $U_{\rm equiv}$ (\r{A}$^2$) & 0.01169(12) & 0.00491(16) \\
				Ni		& $4c / 4e$ & $x$ & 0 & 0 \\
				& & $y$ & 0.70104(15) & 0.70126(21) \\
				& & $z$ & $1/4$ & $1/4$ \\
				& & $U_{\rm equiv}$ (\r{A}$^2$) & 0.01173(55) & 0.00475(64) \\
				Se1		& $8f$ & $x$ & $1/2$ & 0.50806(129) \\
				& & $y$ & 0.08044(8) & 0.07998(12) \\
				& & $z$ & 0.13781(5) & 0.13798(9) \\
				& & $U_{\rm equiv}$ (\r{A}$^2$) & 0.01198(30) & 0.00542(36) \\
				Se2		& $8f$ & $x$ & 0 & -0.0089(13) \\
				& & $y$ & 0.14580(8) & 0.14548(12) \\
				& & $z$ & 0.95076(5) & 0.95108(9) \\
				& & $U_{\rm equiv}$ (\r{A}$^2$) & 0.01029(29) & 0.00495(37) \\
				Se3		& $4c / 4e$ & 0 & 0 & 0 \\
				& & $y$ & 0.32687(12) & 0.32715(16) \\
				& & $z$ & $1/4$ & $1/4$ \\
				& & $U_{\rm equiv}$ (\r{A}$^2$) & 0.01094(41) & 0.00545(49) \\
				& & TW (\%) & $-$ & 53/47 \\
				& & GOF & 1.37 & 2.2 \\
				& & $wR_2$ (\%) & 5.26 & 6.71 \\
				& & $R_1$ (\%) & 2.61 & 3.28 \\
			\end{tabular}
		\end{ruledtabular}
	\end{table}
	
	\subsection{Equilibrium measurements}
	The equilibrium Raman and photoluminescence (PL) spectra of TNS were mainly measured using a Kr/Ar laser, whose wavelength is tuned at 514.5~nm. For the PL measurements, we also employed the 676~nm, 647~nm, 488~nm, and 457~nm lines from the Kr/Ar laser, and 660~nm, 561~nm, and 532~nm lines from several diode-pumped solid-state lasers. The Raman spectra were measured using a triple grating spectrometer equipped with a nitrogen-cooled CCD camera. To calculate the imaginary part of the Raman response function \chipp, we corrected the raw Raman spectra by the Bose factor \cite{Hayes2012}. We measured the PL spectra with the single grating spectrometer and the same CCD camera. For single grating measurements, stray light at the incoming laser wavelength was eliminated using an edge filter.
	
	\subsection{Pump-probe measurements}
	The pump-probe measurements were performed using a Yb-based amplified laser system with a center wavelength of 1030~\microm, a pulse duration of 450~fs, and a repetition rate of 125~kHz. The output from the amplifier was divided into two for the pump and probe beams. We generated the probe pulse at 515~nm from the second-harmonic generation (SHG) of the laser output using a $\beta$-barium borate (BBO) crystal. The probe pulse duration was evaluated as 700~fs by auto-correlation measurements. We employed this SHG pulse for the pump-probe PL measurements. For the pump-probe Raman measurements, the SHG pulse was further tailored at 514.5~nm using a 4$f$ geometry pulse shaper, with a width of 8~\wn and pulse duration of 2.6~ps.
	
	The pump-probe Raman and PL spectra were measured using a single grating spectrometer and a CCD camera. The stray light from the sample was eliminated by edge filters in the pump-probe Raman measurements, whose specs will be shown later. In the pump-probe PL measurements, we employed a notch filter to remove the stray light.
	
	\section{Equilibrium Raman measurements}
	
	\subsection{Raman selection rules}
	We briefly introduce the Raman selection rules of TNS \cite{Hayes2012,Kim2021}. In the high-temperature orthorhombic $Cmcm$ phase above \tc, the Raman active phonons probed in our scattering geometry belong to either \ag or \bg representation of the point group. Their Raman tensors in the $ac$-plane are given by:
	\begin{center}
		\(R_{A_{g}}=
		\left(\begin{array}{cc}
			b_1 & 0 \\
			0 & b_2
		\end{array}\right),
		R_{B_{2g}}=
		\left(\begin{array}{cc}
			0 & b_3 \\
			b_3 & 0 
		\end{array}\right).
		\)
	\end{center}
	Here, $b_1$, $b_2$, and $b_3$ are non-zero matrix elements. The Raman phonon intensity is given by the contraction of the phonon Raman tensor by the incoming $\bf{e_i}$ and outgoing $\bf{e_s}$ photon polarizations:
	\begin{equation}
		I^{\mu}_{e_i,e_s}=|{\bf{e_i}}R_{\mu}{\bf{e_s}}|^{2},
	\end{equation}
	where $\mu$~=~$A_{g}$, $B_{2g}$. In the \raa geometry (the incoming and outgoing photon along the $a$-axis), we have 
	$I^{B_{2g}}_{aa}$~=~0 and $I^{A_{g}}_{aa}$~=~$|b_1| ^2$. In the \rac geometry (incoming and outgoing photon along the $a$-axis and the $c$-axis, respectively), we have $I^{B_{2g}}_{aa}$~=~0 and $I^{A_{g}}_{aa}$~=~$|b_3|^2$.
	Thus, above \tc, the \ag and \bg phonons are probed separately by the \raa and \rac geometries, respectively.
	
	Upon entering the low-temperature monoclinic $C2/c$ phase below \tc, the \ag and \bg representations merge into the $A_g$ representation. All Raman active phonons thus belong to the same representation whose Raman tensors in the $ac$-plane can be written as:
	\begin{center}
		\(R_{A_{g}}=
		\left(\begin{array}{cc}
			b'_1 & b'_3 \\
			b'_3 & b'_2
		\end{array}\right).
		\)
	\end{center}
	where the $b'$ coefficients will generally differ from the $b$ coefficients defined in the orthorhombic phase. Therefore, all Raman phonons have finite Raman intensity in both the \raa and \rac geometry below \tc. The Raman selection rules of phonons can thus serve as symmetry-based indicators of the lattice point group.
	\begin{figure}[t]
		\centering
		\includegraphics[width=\columnwidth]{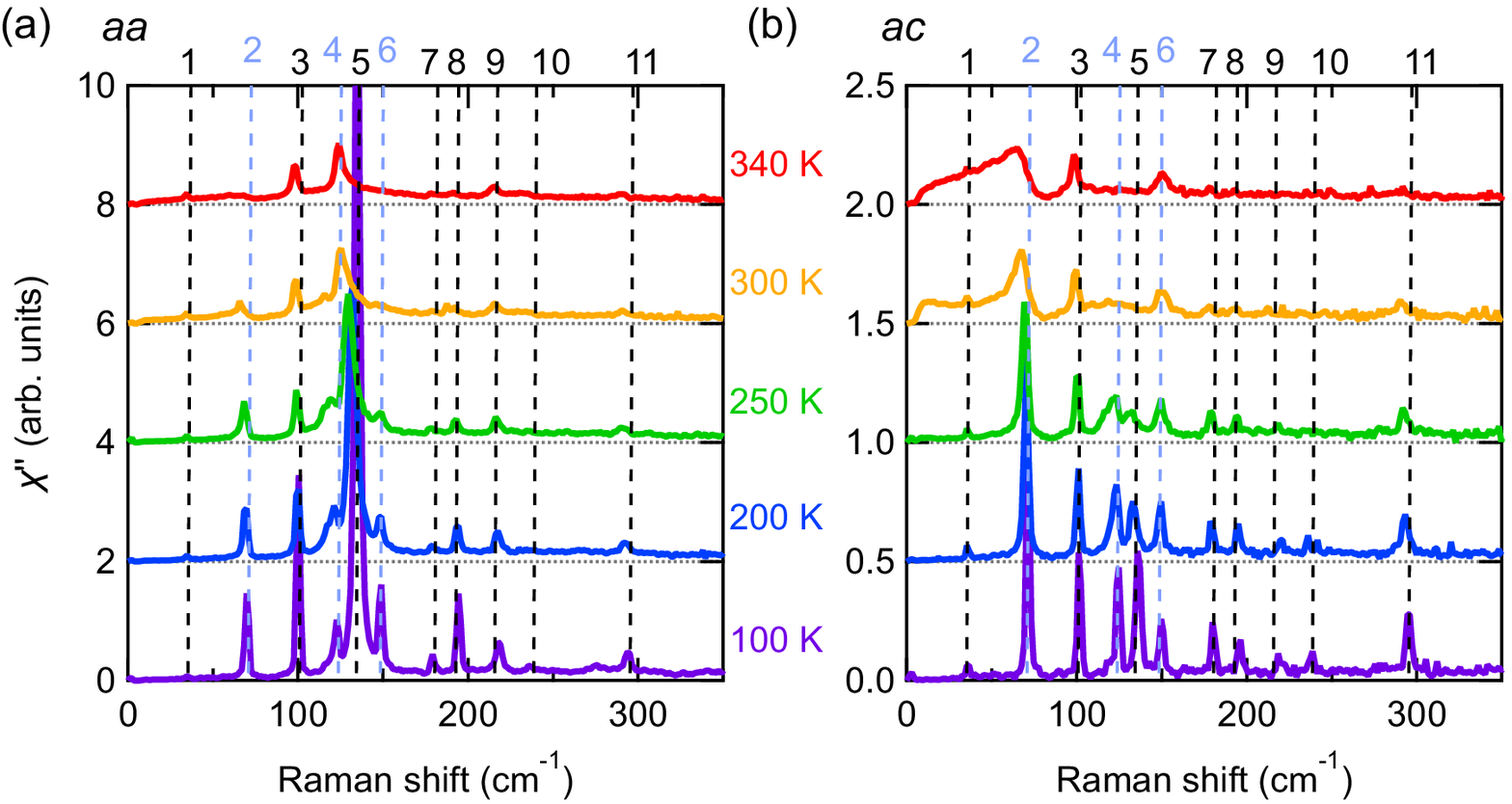}
		\caption{The stokes Raman spectra in equilibrium for the (a) \raa and (b) \rac geometries at selected temperatures. The \ag and \bg phonons defined above \tc are labeled by black and light blue, respectively.}
		\label{figs1}
	\end{figure}
	
	\subsection{Stokes Raman shift data}
	We present the equilibrium Stokes Raman response functions \chipp in Fig. \ref{figs1}. Since the structural symmetry is orthorhombic above \tc, the \raa and \rac geometries probe the \ag and \bg phonons, respectively, following the Raman selection rules. With lowering the temperature, the \bg phonons (mode 2, 4, and 6) emerge in the \raa spectra below \tc, consistent with the anti-Stokes spectra in Fig. 1(a) in the main text.
	
	To evaluate the mode-2 phonon intensities $I_{aa}$ and $I_{ac}$, we integrate the Raman response \chipp from 50~\wn to 80~\wn and subtract the electron continuum at 80~\wn. The obtained intensity ratio between the \raa and \rac geometries is plotted in Fig.~1(c) and increases below \tc.
	
	As mentioned in the main text, the mode-2 phonon displays an asymmetric broadening around \tc with the increase in temperature. Previous reports also observed this behavior and interpreted as a manifestation of the coupling between the electron continuum and the mode-2 phonon \cite{Kim2021,Volkov2021_npj,Ye2021}.
	
	\section{Equilibrium high-energy Raman (PL) measurements}
	\begin{figure}[t]
		\centering
		\includegraphics[width=\columnwidth]{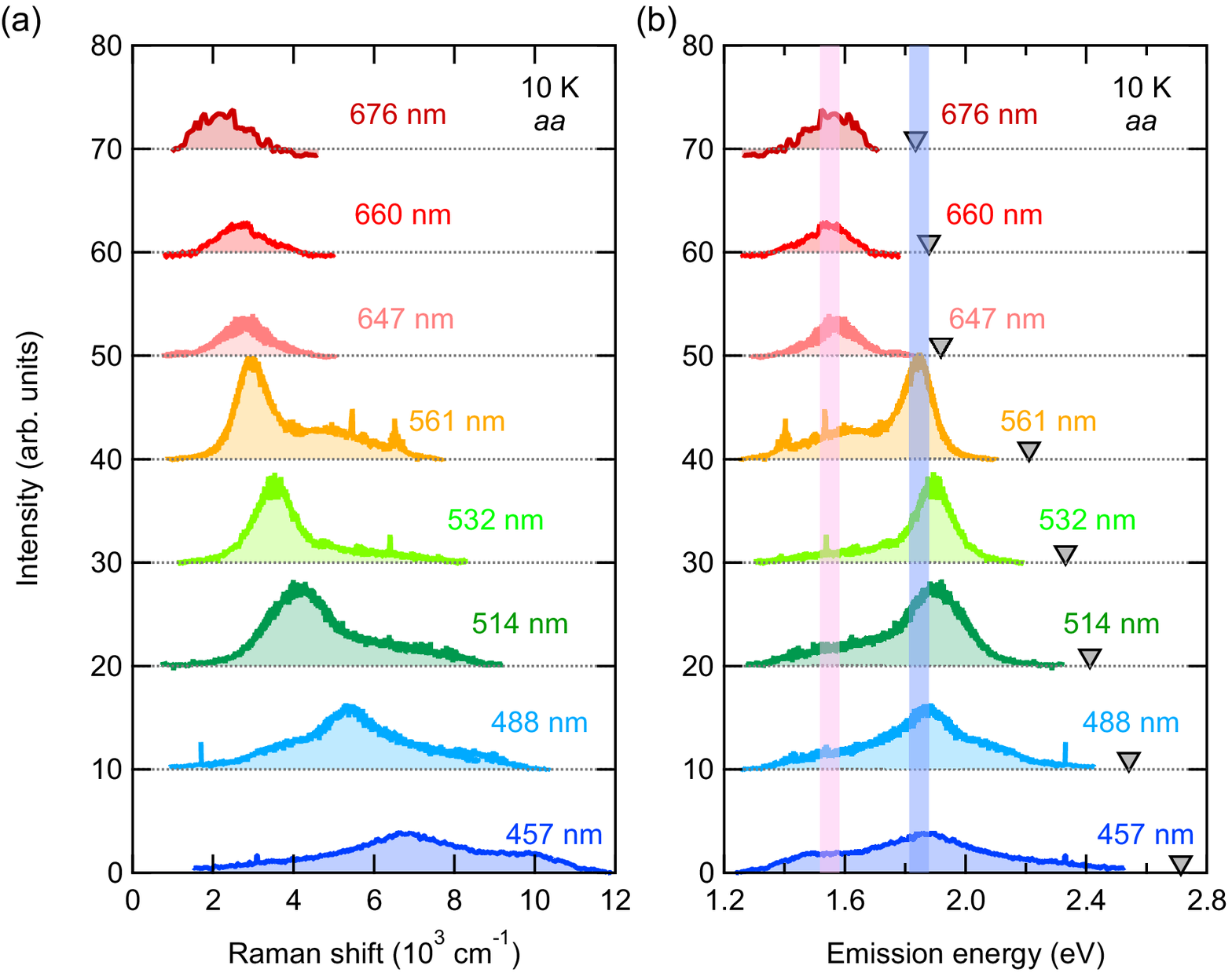}
		\caption{(a) High-energy Raman spectra using various CW-laser wavelengths $\lambda_0$ measured at 10~K for the \raa geometry. (b) The same plot as (a) but as a function of the emitted photon energy. The gray triangles denote the photon energy of each CW-laser wavelength $\lambda_0$.}
		\label{figs2}
	\end{figure}
	
	Figure \ref{figs2}(a) displays the high-energy Raman spectra at 10~K using various wavelengths of the CW lasers $\lambda_0$. We identified two broad peaks for all the measured CW-laser wavelengths. Similar spectra were identified using the CW-laser wavelength of 647~nm and interpreted as Raman signatures of electron-hole excitations across the insulating gap \cite{Volkov2021_npj}. However, the observed peaks locate at different Raman shifts depending on the CW-laser wavelength, evidencing that the observed peaks are not Raman processes. In Fig. \ref{figs2}(b), we plot the spectra as a function of the outgoing photon energy. The peak energies coincide for $457~\text{nm} \leq \lambda_0 \leq 561~\text{nm}$ (shaded light blue) and $676~\text{nm} \leq \lambda_0 \leq 647~\text{nm}$ (shaded light magenta), respectively. This coincidence unambiguously demonstrates that the observed peaks are ascribed to emission processes, in our case, "hot" photoluminescence (PL). 
	
	We depict the general mechanism of a PL process in Fig. 3(b) in the main text \cite{Fox2010}. First, the incoming photon excites electrons from the valence band to a high-energy conduction band (or equivalently from a low-energy valence band to the conduction band). Note that the incoming visible photon energy is much larger than the insulating-gap energy $E_{ex}\simeq$~160~meV reported previously from optical spectroscopy \cite{Lu2017,Larkin2017,Larkin2018,Sugimoto2018}. Secondly, the electrons and holes relax to the bottom of the high-energy conduction band and the top of the valence band (or equivalently to the bottom of the conduction band and the top of the low-energy valence band), respectively. Finally, the recombination of the electrons and holes generates the PL by emitting a photon. In general, radiative processes like PL will compete with much faster non-radiative processes like phonon emission. Hence, the PL intensity will be sizable only when non-radiative processes are sufficiently quenched \cite{Fox2010}. For this reason, we attribute the increase in the PL intensity below \tc to the opening of the large insulating gap, which quenches non-radiative recombination processes involving phonon emission because their energy (typically below 50~meV) is much smaller than the 160~meV gap. We further note that changes in the density of states at the bottom/top of the conduction/valence band and significant changes in optical matrix elements could also contribute to the enhancement of the PL intensity across \tc. 
	
	\begin{figure}[t]
		\centering
		\includegraphics[width=\columnwidth]{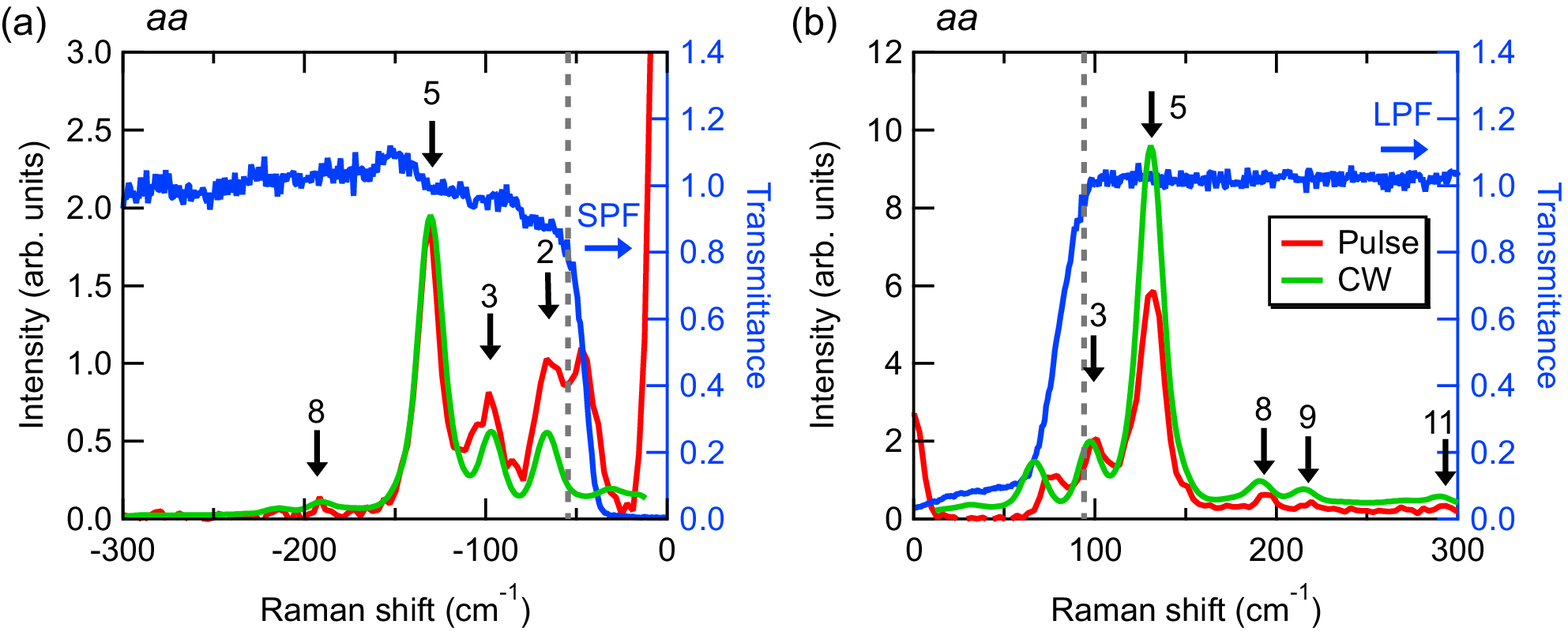}
		\caption{(a) The anti-Stokes Raman spectra in equilibrium for the \raa geometry at 150~K (red). The green curve is the CW anti-Stokes spectra convoluted with the spectral width of the pulse laser. The transmittance of the short pass filter (SPF) is plotted on the right axis. (b) The same plot as (a) but for the Stokes Raman spectra. The transmittance of the long pass filter (LPF) is plotted on the right axis. The gray vertical dashed lines denote the cutoff of the filters.}
		\label{figs3}
	\end{figure}
	
	\section {Pump-probe Raman measurements}
	\subsection{Spec of the edge filters}
	In pump-probe Raman measurements, we have to be careful about the transmittance of the edge filters because it blocks part of the Raman signal. Figure \ref{figs3} compares the equilibrium anti-Stokes/Stokes Raman spectra using the pulsed laser (the red curves) and those using the CW laser convoluted with a Gaussian whose full width at half maximum is 9~\wn (the green curves). The transmittance of the short/long pass filters is also shown by the blue curves on the right axis. Both the anti-Stokes and Stokes spectra using the pulsed laser match reasonably well those of the convoluted CW spectra. Note that the filter on the anti-Stokes has a cutoff at a lower absolute Raman shift than on the Stokes side, allowing the mode-2 phonon to be more easily resolved in this geometry. Hence, we employed the anti-Stokes spectra to investigate the mode-2 phonon. In the anti-Stokes spectra, we found a peak around $-47$~\wn, which does not correspond to any phonons in the CW spectra. This peak is most likely due to the leakage from stray light imperfectly blocked by the filter.
	
	\begin{figure}[t]
		\centering
		\includegraphics[width=\columnwidth]{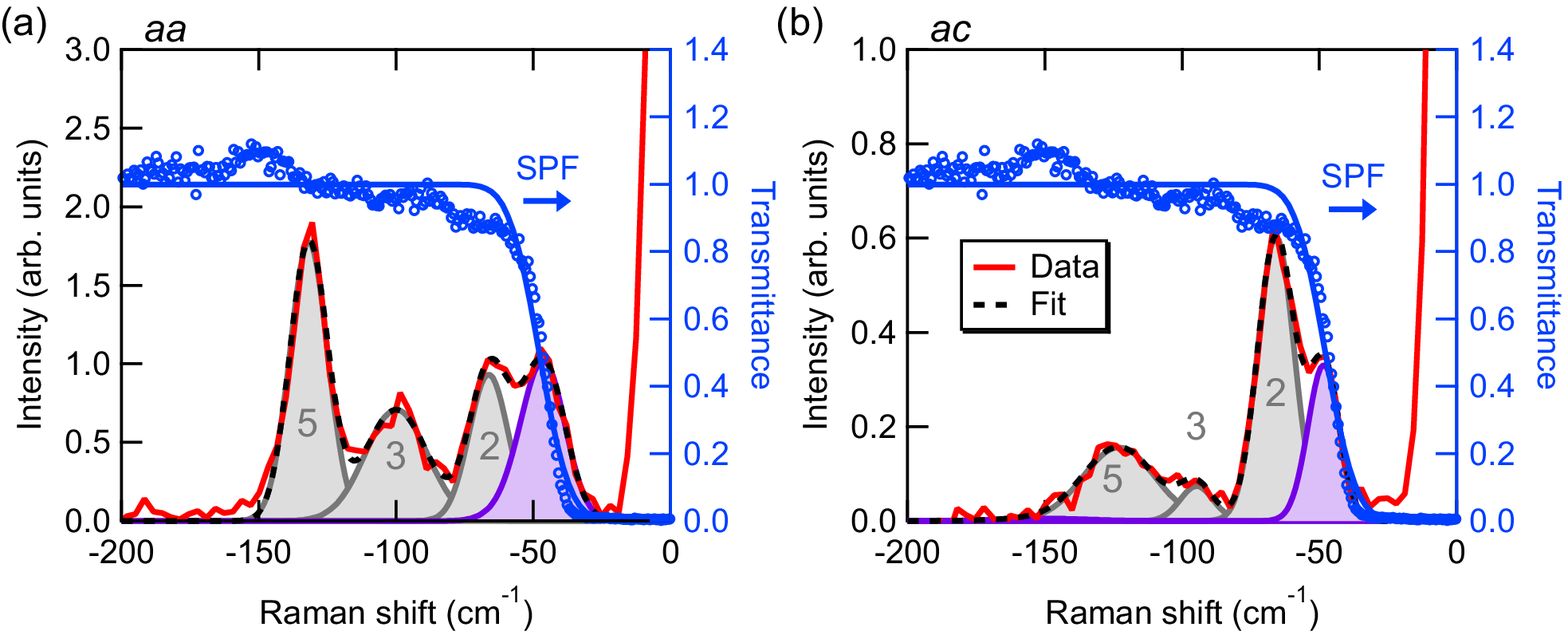}
		\caption{(a) The fitting results of the anti-Stokes spectra at 150~K in equilibrium for (a) the \raa and (b) \rac geometries, respectively (left axis). The contributions from the mode 2, 3, and 5 phonons are also shown by the gray curves. The purple curve is the fit to the leakage signal with a Gaussian shape multiplied by an error function. The transmittance of the SPF is presented by the blue open circles (right axis). The blue curve shows the fit to the SPF transmittance with an error function.}
		\label{figs4}
	\end{figure}
	
	\begin{figure*}[t]
		\centering
		\includegraphics[width=\linewidth]{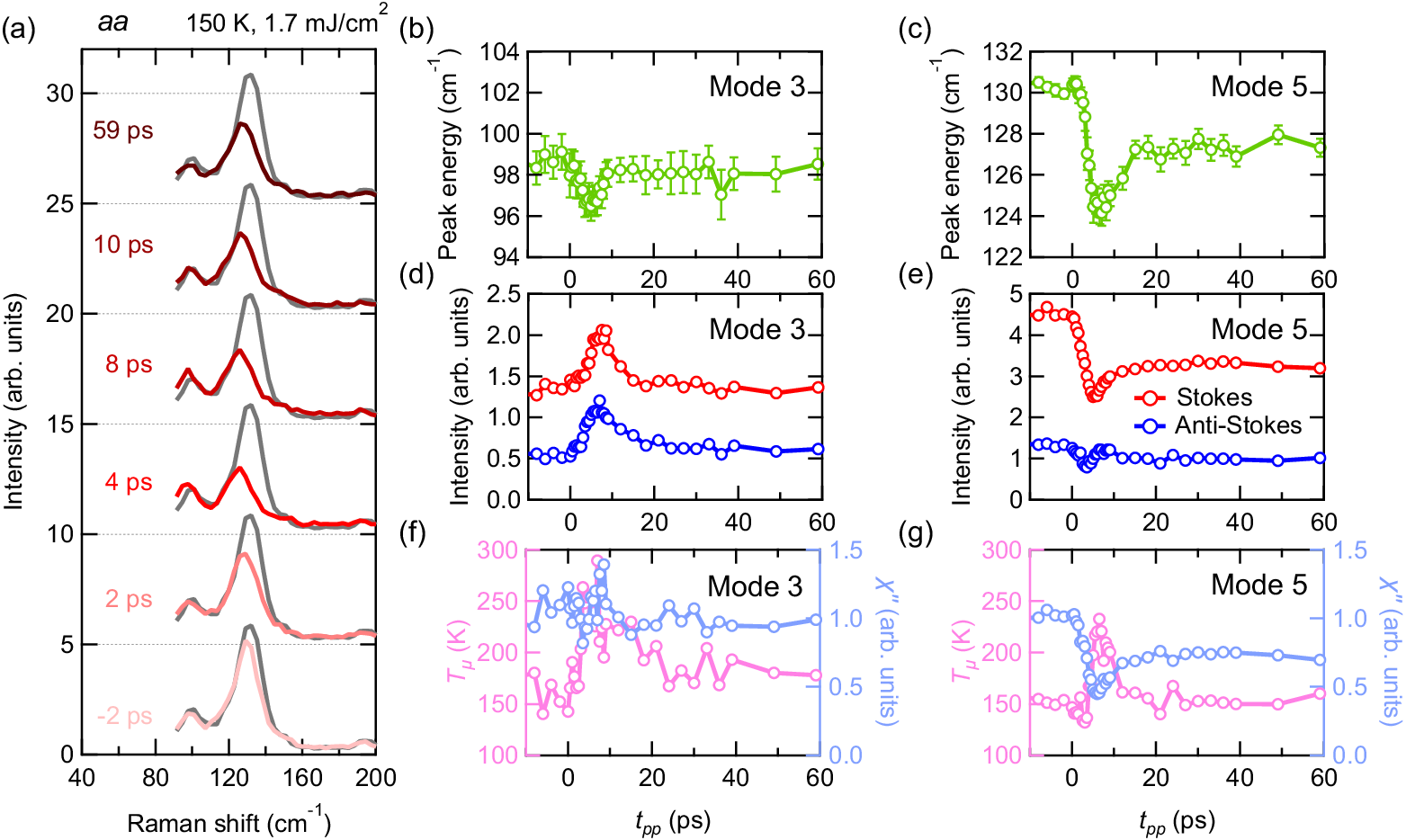}
		\caption{(a) Transient Stokes Raman spectra for the \raa geometry at 150~K with the pump fluence of 1.7 \mjc. (b,c) The phonon peak energy as a function of the pump-probe delay time \tpp for (b) mode-3 and (c) mode-5 phonons, respectively. The error bars are fitting errors. (d,e) The phonon peak intensity as a function of \tpp for (d) mode-3 and (e) mode-5 phonons, respectively. (f,g) The time-evolution of the phonon temperature \tmu (left axis, magenta) and Raman response function \chipp (right axis, light blue) for (f) mode-3 and (g) mode-5 phonons, respectively.}
		\label{figs5}
	\end{figure*}
	
	\subsection{Evaluation of the phonon intensities by fitting}
	We evaluated the intensities of the mode 2, 3, and 5 phonons by fitting the Raman spectra with Gaussian shapes because the spectral width of the probe pulse is larger than that of the phonons. We included the contributions from the stray light leakage signal using a Gaussian shape multiplied by an error function fitted to the SPF transmittance. The fitting curves to the anti-Stokes spectra at 150~K in equilibrium are displayed in Fig. \ref{figs4}. We removed the leakage signal from the anti-Stokes spectra, and showed the obtained spectra in Fig.~2(a) in the main text. Using the obtained mode-2 phonon intensities for the \raa and \rac geometries, we computed their ratio $I_{aa}/I_{ac}$, which is presented in Fig.~2(b) in the main text. 
	
	\subsection{Stokes Raman spectra for the \raa geometry}
	Figure \ref{figs5}(a) shows the transient Stokes Raman spectra for the \raa geometry at 150~K with the incident pump fluence of 1.7~\mjc, which was measured with the same excitation condition as the anti-Stokes spectra in Fig.~2(a) in the main text. Similar to the anti-Stokes spectra, the mode-3 phonon intensity increases while the mode-5 phonon intensity decreases after photoexcitation. We evaluated each phonon intensity assuming a Gaussian lineshape. Figures~\ref{figs5}(b) and (c) display the time evolution of the energy of mode-3 and mode-5 phonons, respectively. We also present the time evolution of the intensity of the mode-3 and mode-5 phonons in Figs. ~\ref{figs5}(d) and (e), respectively. Of particular interest is the sizable softening of mode-5 phonon, which sustains even up to 59~ps.
	As mentioned in the main text, the out-of-equilibrium Raman phonon intensity will depend on both the effective mode temperature (\tmu) and the phonon Raman response function (\chipp). The two effects can be separated using the Stokes and anti-Stokes spectra \cite{Hayes2012}. According to the fluctuation-dissipation theorem, the Stokes (S) and anti-Stokes (AS) intensities are written as
	\begin{align}
		I_{\text{S}}(\omega) &= (n(\omega)+1)\chi^{\prime\prime}(\omega) \label{eq1}\\
		I_{\text{AS}}(\omega) &= n(\omega)\chi^{\prime\prime}(\omega). \label{eq2}
	\end{align}
	Here, the $n(\omega)$ is the occupation number and given by
	\begin{align}
		n(\omega) = \frac{1}{\exp(\hbar\omega/k_BT_{\mu})-1},
		\label{eq3}
	\end{align}
	where $\hbar$ is the Planck constant divided by $2\pi$ and $k_B$ is the Boltzmann constant. Combining Eqs. \eqref{eq1}-\eqref{eq3}, the temperature and Raman response function of a phonon are given by
	\begin{align}
		T_{\mu}(\omega)& = \frac{\hbar\omega}{k_B\ln({I_{\text{S}}(\omega)/I_{\text{AS}}(\omega)})}, 	\label{eq4}\\
		\chi^{\prime\prime}(\omega) &= I_{\text{S}}-I_{\text{AS}}. \label{eq5}
	\end{align}
	
	In Figs. \ref{figs5}(f) and (g), we present the obtained effective temperature \tmu and Raman response function \chipp of mode-3 and mode-5 phonons using Eqs. \eqref{eq4} and \eqref{eq5}. For the mode-3 phonon, the temperature \tmu increases upon photoexcitation, while the Raman response function \chipp does not show significant change. For the mode-5 phonon, \tmu increases after photoexcitation and relaxes to its equilibrium value around \tpp~=~10~ps. On the contrary, the \chipp of the mode-5 phonon decreases and reaches the metastable value at \tpp~=~10~ps, which sustains even up to \tpp~=~59~ps, tracking the behavior of its energy. This indicates that the mode-5 phonon reaches a nonequilibrium state, which is distinct from the equilibrium state and cannot be described by a simple temperature increase, possibly pointing toward a subtle metastable structural distortion. 
	
	\begin{figure*}[t]
		\centering
		\includegraphics[width=\linewidth]{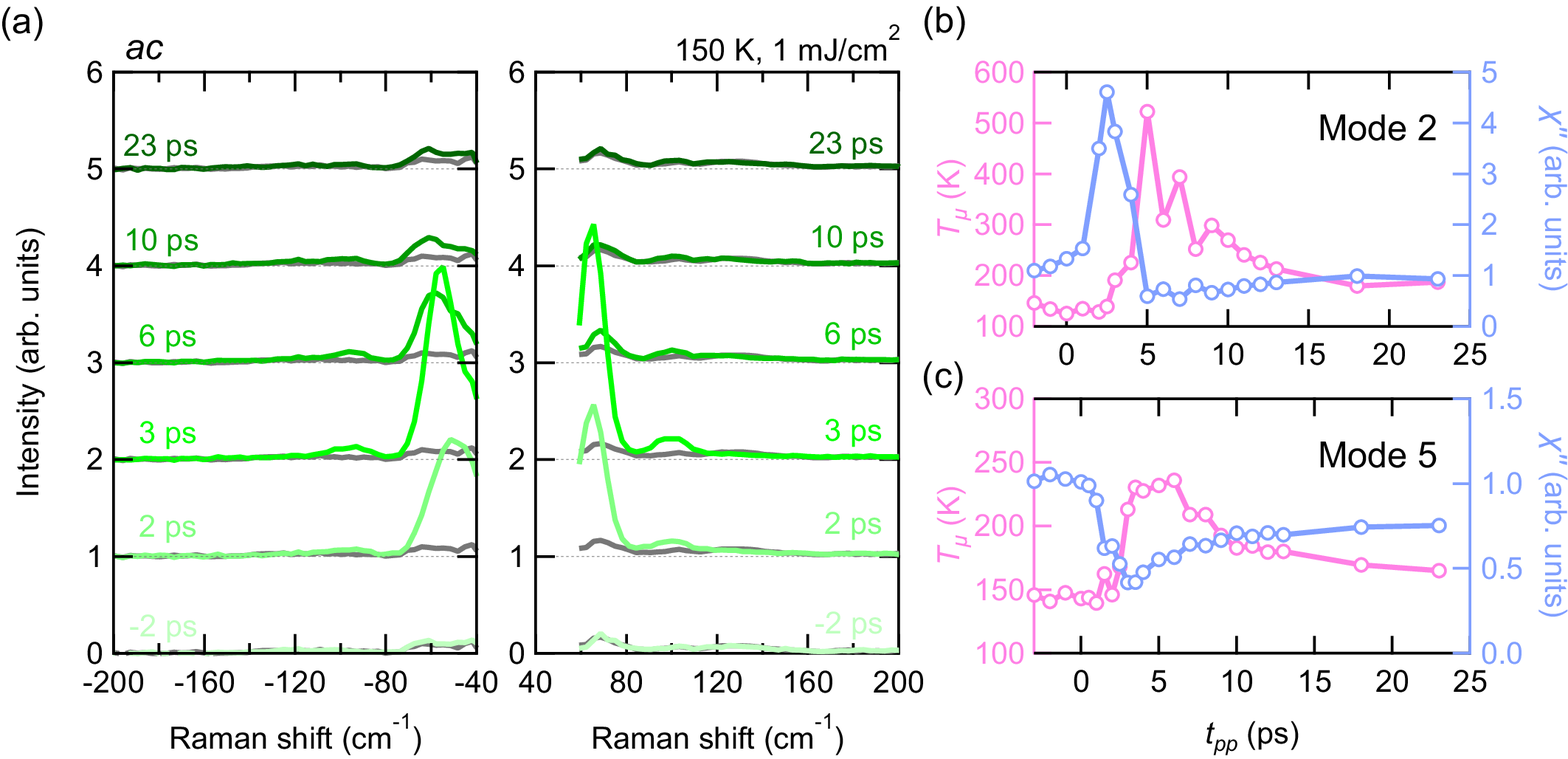}
		\caption{(a) Transient Stokes (left) and anti-Stokes (right) Raman spectra for the \rac geometry at 150~K with the pump fluence of 1~\mjc. (b) Time evolution of the mode-2 phonon temperature \tmu (left axis, magenta) and Raman response function \chipp (right axis, light blue). (c) Same plot as (b) but for the mode-5 phonon.}
		\label{figs6}
	\end{figure*}
	
	\subsection{Additional Raman spectra for the \rac geometry}
	Though the mode-2 phonon could not be observed experimentally in the \raa Stokes spectra in the initial measurements, we could observe it in a separate measurement on a cleaner sample surface in the \rac geometry where the mode-2 intensity is more significant than that in the \raa geometry. This is shown in the right panel of Fig. \ref{figs6}(a). Here, the incoming pump fluence is set to 1~\mjc, lower than for the measurements reported in the main text. The anti-Stokes and Stokes spectra are extended to Raman shifts below the respective filter cutoff, by dividing the raw spectra with the filter transmission. It should also be noted that for these measurements, the spectral and time width of the probe pulse are 9~\wn and 2.3~ps, respectively, slightly different from the one reported in the main text. Combining the \rac anti-Stokes spectra performed in the same excitation condition, we can evaluate the transient temperature and the Raman response function of the mode-2 phonon as presented in Fig. \ref{figs6}(b). The Raman response function \chipp increases instantaneously and relaxes to the equilibrium value after \tpp~=~5~ps. On the other hand, the temperature \tmu takes the maximum value around \tpp~=~5~ps and reaches a metastable value after \tpp~=~18~ps. For comparison, we represent \tmu and \chipp for the mode-5 phonon using the \raa geometry under the same excitation condition in Fig.~\ref{figs6}(c). The results are very similar to the one for the pump fluence of 1.7~\mjc in Fig.~\ref{figs5}(c). Right after photoexcitation, both the changes in \tmu and \chipp for the mode-2 phonon are larger than those for the mode-5 phonon. This provides further evidence that the mode-2 phonon strongly couples to (photoexcited)-electrons consistently with previous equilibrium Raman measurements \cite{Kim2021,Volkov2021_npj,Ye2021} and the frequency-resolved pump-probe ARPES experiment \cite{Suzuki2021}. In addition, it is notable that at least 8~ps after photoexcitation, both the mode-2 and mode-5 phonon temperatures are lower than \tc, indicating that a simple temperature increase cannot account for the nonequilibrium state with the reduced PL intensity found after 3~ps (Fig.~4(c) in the main text).
	
	
	\bibliography{TNS.bib}
	\bibliographystyle{apsrev4-1}